\documentclass{article}
\usepackage{graphicx} % Required for inserting images
\usepackage{amsmath}
\usepackage{amssymb}
\usepackage{amsthm}
\usepackage{thmtools}
\usepackage{thm-restate}
\usepackage{bm}
\usepackage{graphicx}
\usepackage{caption}
\usepackage[margin = 1in]{geometry}
\usepackage{natbib}
\usepackage[normalem]{ulem}
\setcitestyle{numbers,square}
\usepackage{color}
\usepackage[dvipsnames]{xcolor}
\usepackage{media9}
\usepackage{subcaption}
\usepackage{comment}
\usepackage{enumitem}
\usepackage[colorlinks,urlcolor=blue,citecolor=ForestGreen,linkcolor=ForestGreen]{hyperref}
\usepackage[nameinlink,capitalize,nosort]{cleveref}
\crefname{claim}{Claim}{Claims}
\usepackage{palatino}

\usepackage{algorithm}
\usepackage[noend]{algpseudocode}

\usepackage[colorinlistoftodos]{todonotes}
\usepackage[most]{tcolorbox}

\usepackage[utf8]{inputenc}

\definecolor{mygreen}{rgb}{0,0.65,0.66}

\newcommand{\vol}{\text{vol}}
\newcommand{\con}{\text{con}}

\newcommand{\mcP}{\mathcal{P}}

\newcommand{\stc}{\text{stc}}

\graphicspath{{imgs/}}

\makeatletter
\def\mathcolor#1#{\@mathcolor{#1}}
\def\@mathcolor#1#2#3{%
  \protect\leavevmode
  \begingroup
    \color#1{#2}#3%
  \endgroup
}
\makeatother

\newcommand*{\vsepfbox}[1]{%
  \begingroup
    \sbox0{\fbox{#1}}%
    \setlength{\fboxrule}{0pt}%
    \mbox{\kern-\fboxsep\fbox{\unhbox0}\kern-\fboxsep}%
  \endgroup
}

\theoremstyle{plain} \numberwithin{equation}{section}
\newtheorem{theorem}{Theorem}[section]
\numberwithin{theorem}{section}

\newtheorem{lemma}[theorem]{Lemma}

\newtheorem{definition}[theorem]{Definition}

\theoremstyle{definition}

\usepackage{thmtools} 
\usepackage{thm-restate}

\makeatletter
\newcommand{\subalign}[1]{%
  \vcenter{%
    \Let@ \restore@math@cr \default@tag
    \baselineskip\fontdimen10 \scriptfont\tw@
    \advance\baselineskip\fontdimen12 \scriptfont\tw@
    \lineskip\thr@@\fontdimen8 \scriptfont\thr@@
    \lineskiplimit\lineskip
    \ialign{\hfil$\m@th\scriptstyle##$&$\m@th\scriptstyle{}##$\hfil\crcr
      #1\crcr
    }%
  }%
}
\makeatother

\newcounter{note}[section]

\newcommand{\eps}{\epsilon}

\title{Flows in High Degree Places:\\ Sublinear Spanning Tree Congestion}
\author{%
   \begin{tabular}{cc}
     D Ellis Hershkowitz\thanks{Supported in part by NSF grant CCF-2403236 and NSF CAREER Award CCF-2540331.} & \hspace{0.5cm} Richard Z Huang\footnotemark[1]
     \\\\
     \multicolumn{2}{c}{Brown University}
   \end{tabular}
 }
\date{}

\begin{document}

\maketitle
 
\begin{abstract}
    The congestion of a spanning tree $T$ of graph $G$ is the maximum congestion of an edge of $T$ if each edge of $G$ is routed on $T$. 
    Prior work showed that obtaining a minimum congestion spanning tree is NP-hard to approximate within $6/5$, yet the only known approximation algorithm is a trivial $n/2$-approximation from returning any spanning tree of the graph. 
    We give a simple poly-time $\tilde{O}(n^{6/7})$-approximation for minimum congestion spanning tree on $n$-vertex graphs. We obtain this result via a $\tilde{O}(m^{2/3})$-approximation, where $m$ is the number of edges. Our algorithm is based on a new \textit{flow-or-degree decomposition} which deletes a small number of edges so that each remaining component either has low degree or admits a low congestion flow. We also show that, assuming the Small Set Expansion Hypothesis, it is NP-hard to $O(1)$-approximate this problem, even on graphs of constant maximum degree.
\end{abstract}
\pagenumbering{gobble}

\newpage
\pagenumbering{arabic}
\setcounter{page}{1}

\section{Introduction}
Spanning tree congestion (STC) is a measure of the routing quality of a tree. Roughly, it is defined as the maximum congestion of any edge if we route each edge of the graph in the tree. Formally, it is as follows.
\begin{definition}[Spanning Tree Congestion]
The congestion of a spanning tree $T$ of a graph $G=(V,E)$ is
\begin{align*}
    \con_G(T) := \max_{e\in E(T)}\left|\left\{\{u,v\}\in E : \text{the unique $u$--$v$ path in $T$ contains $e$}\right\}\right|.
\end{align*}
\end{definition}
\noindent We refer to the minimum congestion of a spanning tree of $G$ as the spanning tree congestion of $G$ and denote it by $\stc(G)$.

To see why spanning tree congestion is a useful notion for understanding routing, consider the following example of \cite{law2010spanning}. Suppose we have a graph $G$ representing the subway system of a city. Furthermore, suppose the city needs to close as many routes as possible for maintenance without disconnecting the network where, ideally, the extra congestion incurred by closures is minimized. Then, the optimal solution for such a problem is the spanning tree of the subway network of congestion $\stc(G)$.

 Absolute bounds on the spanning tree congestion of a graph have been extensively studied over the past several decades. In particular, spanning tree congestion was introduced in \cite{ostrovskii2004minimal}.\footnote{\cite{simonson1987variation} also studied spanning tree congestion under a different name.} Since then, there have been numerous works on its combinatorics \cite{ostrovskii2004minimal,kozawa2009spanning,law2009spanning,ostrovskii2011minimum,kozawa2011spanning,law2011spanning,bodlaender2011spanning,law2013spanning}. Notably, \cite{chandran2018sqrtmn} showed that $n$-node, $m$-edge graphs always have a (poly-time computable) spanning tree of congestion $\tilde{O}(\sqrt{mn})$.\footnote{The notation $\tilde{O}$ suppresses factors polylogarithmic in $n$.} Conversely, a simple lower bound shows $\stc(G)\geq m/n$ \cite{otachi2011designing}. For a more comprehensive history, we refer the reader to \cite{otachi2020survey}. %There has also been  its computational and parameterized complexity \cite{otachi2010complexity,okamoto2011hardness,bodlaender2012parameterized,luu2025better,lampis2024parameterized,atalig2026two,otachi2026spanning}

However, our understanding of poly-time approximation algorithms for minimum congestion spanning trees is much less developed. In particular, the spanning tree congestion problem asks, given a connected graph $G$, for a minimum congestion spanning tree $T$ of $G$. For many years the only known poly-time approximation for this problem was a trivial $n/2$-approximation obtained by returning \textit{any} spanning tree of the graph. On the other hand, the previously best known hardness of approximation lower bound for poly-time algorithms was a remarkably smaller $6/5$~\cite{luu2025better}. Motivated by the limited progress on approximation algorithms, Otachi~\cite{otachi2020survey} posed the following open question in his survey on spanning tree congestion:
\begin{quote}\centering
    \textit{Is there a poly-time $o(n)$-approximation for spanning tree congestion?}
\end{quote}
\noindent Recently, \cite{kolman2025deltaapx} made partial progress on this question by giving a poly-time $\tilde{O}(\Delta_G)$-approximation in graphs of maximum degree $\Delta_G$. However, this does not give a nontrivial approximation for general graphs.

The lack of progress on approximation algorithms for this problem is likely explained by several obstacles. First, a subgraph can have STC arbitrarily larger than its parent graph---for example, a graph composed of a constant-degree expander along with a vertex attached to all vertices has STC $O(1)$ whereas the expander subgraph has STC  $\Omega(n)$~\cite{kolman2025deltaapx}. As such, using recursive algorithms for STC is rather challenging. Similarly, contraction of connected components is not clearly a safe operation from an approximation perspective. Even worse, the natural LP relaxation has an integrality gap of $\Omega(n)$; see \Cref{sec:IG}.

\subsection{Our Contributions}
We give a simple poly-time $\Tilde{O}\left(n^{6/7}\right) = o(n)$-approximation for spanning tree congestion.\footnote{Our analysis can be improved to prove an approximation guarantee of $\Tilde{O}\left(n^{3/4}\right)$. For ease of presentation we omit these details.}
\begin{tcolorbox}
    \begin{restatable}[Sublinear approximation]{theorem}{thmsublinear}\label{thm:main}
        There is a poly-time algorithm that, given a graph $G$, returns a spanning tree $T$ with congestion at most $\stc(G) \cdot \Tilde{O}\left(n^{6/7}\right)$.
    \end{restatable}
\end{tcolorbox}
\noindent This considerably improves on the previous trivial $n/2$-approximation for general graphs.

We prove our $o(n)$-approximation result by giving an $o(m)$-approximation. In particular, it is a simple exercise (see \Cref{subsec:ntom}) to see that the aforementioned $\Tilde{O}(\sqrt{mn})$ congestion spanning trees and the lower bound of $\Omega(m/n)$ imply that an $m^\eps$-approximation for sufficiently small $\eps$ gives an $n^\delta$-approximation for $\delta < 1$ depending only on $\eps$. We therefore give the following $o(m)$-approximation.
\begin{tcolorbox}
    \begin{restatable}{theorem}{thmmain}\label{thm:sparse}
        There is a poly-time algorithm that, given a graph $G$, returns a spanning tree $T$ with congestion at most $\stc(G) \cdot \Tilde{O}\left (m^{2/3}\right )$.
    \end{restatable}
\end{tcolorbox}
\noindent 
Complementing our approximation algorithms, we show that, under the Small Set Expansion Hypothesis (a standard complexity assumption), it is NP-hard to approximate minimum spanning tree congestion within any constant factor.
\begin{tcolorbox}
    \begin{restatable}[SSEH Hardness]{theorem}{lowerbound}\label{thm:hardness}
        Assuming the Small Set Expansion Hypothesis, there is a constant $D>0$ such that it is NP-hard to approximate spanning tree congestion within any constant factor, even on graphs with maximum degree at most $D$.
    \end{restatable}
\end{tcolorbox}

\noindent  This considerably improves on the previous $6/5$ hardness of approximation \cite{luu2025better}.

\subsection{Overview and Intuition}
Before moving on to formal details, we give a high-level overview of both our $o(m)$-approximation and our $\omega(1)$ hardness of approximation.

\subsubsection{$o(m)$-Approximation Overview and Intuition}

At a high level, our algorithm for graph $G= (V,E)$ is very simple. First, we apply what we call a \textit{flow-or-degree decomposition}, a small set of edges $I \subseteq E$ such that each connected component of $G-I$ either consists of vertices with low degree or admits a low node congestion flow from each vertex to a designated root.\footnote{For these flows, node congestion is the maximum flow through a node other than the root.}
\begin{itemize}
    \item For the low-degree components, we apply the $\tilde{O}(\Delta_G)$-approximation of \cite{kolman2025deltaapx} to compute a spanning tree on each such component. One subtlety is that ``low degree'' means low degree in $G$, not the graph induced by the connected component, and our guarantee requires a minor refinement of the analysis of \cite{kolman2025deltaapx}---see \Cref{subsec:degapx}.
    \item For the low congestion flow components, we take this flow and then round it to a spanning tree on the component of similar congestion, up to polylogarithmic factors, using confluent flows---see \Cref{subsec:confluent}. This idea was previously used in the context of spanning tree congestion by \cite{chandran2018sqrtmn}.
\end{itemize}
As $I$ consists of a small number of edges, it is then safe to connect up the resulting trees from the connected components of $G-I$ arbitrarily (since each such arbitrary connection is only congested by edges of $I$).

Roughly, our flow-or-degree decomposition $I$ deletes $O(m^{2/3})$ edges and guarantees that each component $C$ of $G-I$ either has maximum degree in $G$ at most $O(m^{2/3})$ or admits a flow in $G[C]$ to a root $r\in C$ in which every vertex $v\in C\setminus\{r\}$ sends $\deg_G(v)$ flow, with node congestion at most $O(m^{2/3})\cdot\stc(G)$. See \Cref{lemma:decomp}. Thus, it only remains to discuss how we compute our flow-or-degree decomposition.

\paragraph{Flow-Or-Degree Decomposition.} One can make significant progress towards our decomposition by simply taking a cut separating all high degree nodes (i.e.\ a multiway cut). In particular, the set $H$ of nodes of degree at least $m^{2/3}$ certainly has size at most $O(m^{1/3})$. Thus, if we compute a cut separating all such vertices, we end up with at most one vertex in each connected component of degree at least $m^{2/3}$. Furthermore, it is a simple exercise to see that such a multiway cut can be computed with cost at most $O(m^{1/3})\cdot\stc(G)$---see \Cref{lemma:multiwaycut}. The remaining issue is, of course, that the components with a high degree node may not admit a low congestion flow.

To address this, we use the idea of a cutmatch. In particular, it is well-known that max-flow-min-cut duality implies that, given candidate sources and sinks for a (single commodity) flow, the subset of sources which cannot send flow with low congestion can be certified as such by a sparse cut. This has been used to great effect in, for example, the expander decompositions literature \cite{khandekar2009graph,saranurak2019expander,haeupler2023maximum,haeupler2022hop} which formalizes the low congestion flow and sparse cut as a ``cutmatch''. Formalizing the appropriate sense of cutmatch for us gives us a small collection of low degree nodes whose boundary edges we can delete so that the remaining component containing the high degree node admits a low congestion degree-to-root flow. See \Cref{subsec:cutmatch}. Taking this cut in each component gives our final decomposition. %Thus, the last step of our algorithm is to apply the cut of a cutmatch to each component with a high degree node to ensure that all nodes which remain in a component with a high degree node have a low congestion flow to the high degree node.

\begin{figure}[H]
    \centering
    \begin{subfigure}[t]{0.5\textwidth}
        \centering
        \includegraphics[width=.79\linewidth]{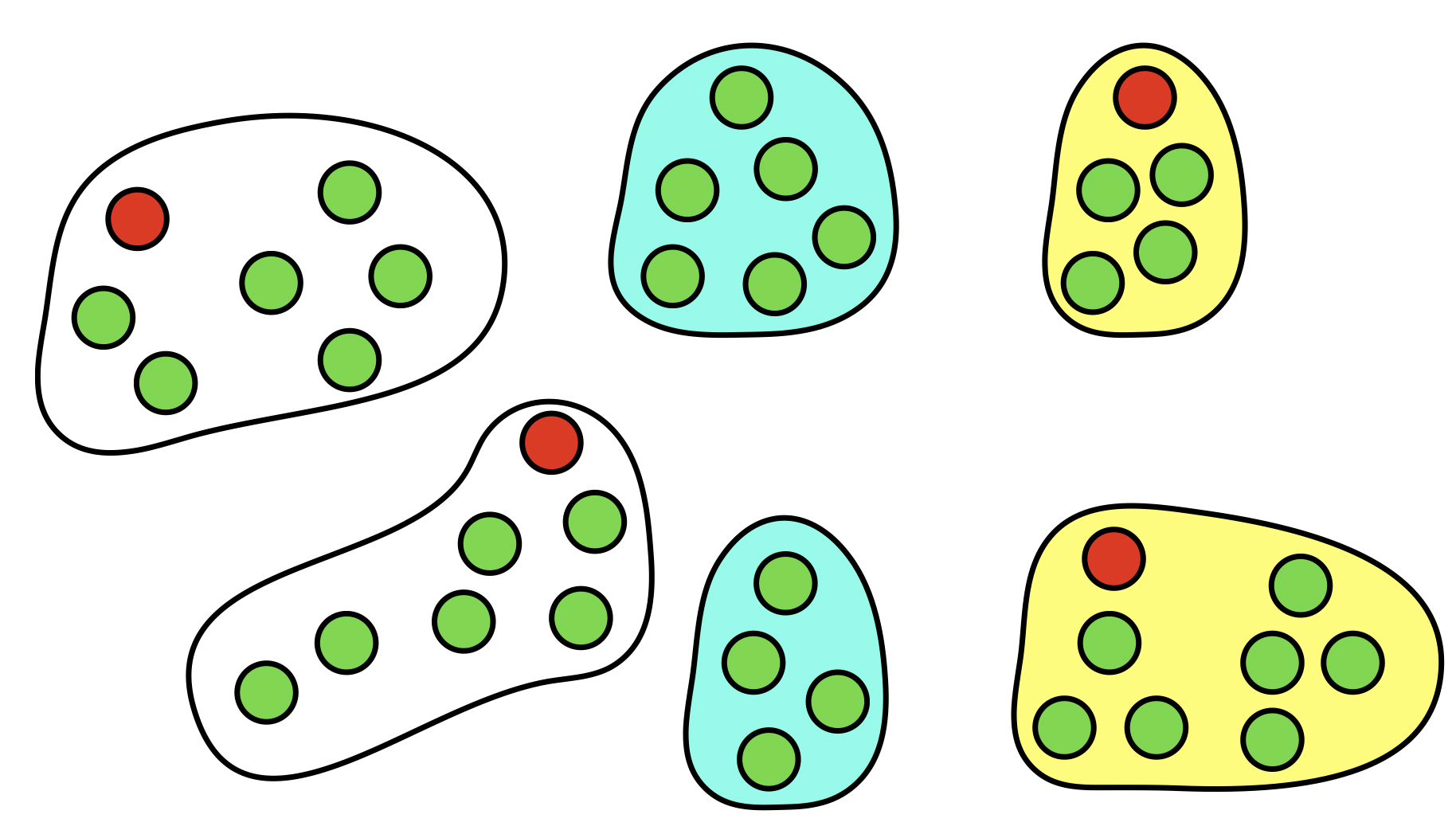}
        \caption{Applying a multiway cut.}
        \label{subfig:decomp1}
    \end{subfigure}%
    ~ 
    \begin{subfigure}[t]{0.5\textwidth}
        \centering
        \includegraphics[width=.79\linewidth]{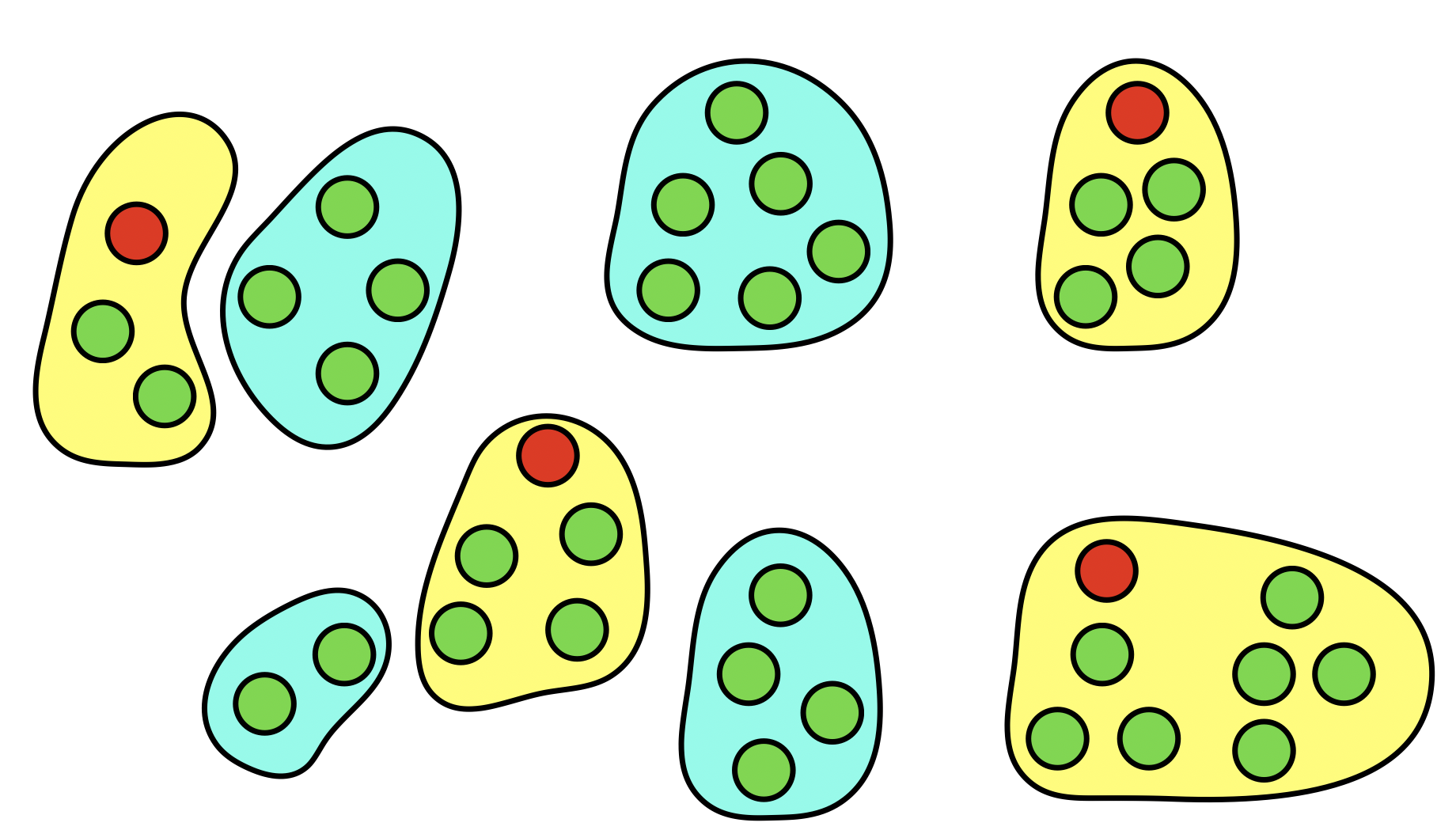}
        \caption{Applying cutmatch.}
                \label{subfig:decomp2}
    \end{subfigure}
    \caption{Computing a flow-or-degree decomposition. High-degree vertices are red and low-degree vertices are green. \Cref{subfig:decomp1}: we compute a multiway cut to separate the high-degree vertices so that each component has at most one high-degree vertex. Yellow components admit a low congestion flow and blue components are low degree. \Cref{subfig:decomp2}: we apply cutmatch to the high-degree components that don't yet admit a low congestion flow, further decomposing them into blue and yellow components. }
    \label{fig:decomp}
\end{figure}

In summary, our algorithm for a flow-or-degree decomposition is roughly as follows. First, take a multiway cut of high degree nodes. Then, on each component with a high degree node, use a cutmatch to separate the nodes that cannot route to that node with low congestion. See \Cref{fig:decomp} for an illustration of the decomposition. One minor caveat here is that, in reality, we guess the target congestion $c=m^{2/3}\cdot\stc(G)$, using degree threshold $c/m^{1/3}$ for each guess.

\subsubsection{$\omega(1)$-Hardness of Approximation Overview and Intuition}
We make use of a particular form of the Small Set Expansion Hypothesis---introduced in \cite{raghavendra2012reductions}---which informally states that, given a graph $G=(V,E)$, it is NP-hard to distinguish between the case in which there is a partition of $V$ such that each part is small and there are few edges crossing the partition (the ``YES'' case) and the case in which every balanced subset of $V$ has many edges leaving it (the ``NO'' case).

The YES case almost immediately gives upper bounds on the spanning tree congestion, and the NO case almost immediately gives lower bounds on the spanning tree congestion. In particular, in the YES case, the tree obtained by taking a spanning tree of each part and then connecting these trees arbitrarily has congestion at most the sum of the number of edges within a part and the number of crossing edges, which are both small. In the NO case, there is a lower bound on the congestion of the minimum congestion tree. In particular, the components resulting from removing the centroid can be agglomerated into large sets. Since we are in the NO case, these balanced sets must have many edges leaving them, all of which congest edges incident to the centroid.\footnote{The centroid of a tree is a vertex whose removal leaves connected components all of size at most $n/2$.} Thus, as long as the centroid has low degree, we get large spanning tree congestion. The caveat to getting this to work is that it requires the centroid to have low degree. We achieve this with one intermediate step in which we transform the graph into a constant-degree graph by, roughly, replacing each vertex with a constant-degree expander.

It follows that any constant-factor approximation for spanning tree congestion would allow us to differentiate between the YES and NO cases above, which is NP-hard assuming the Small Set Expansion Hypothesis.

\subsection{Related Work}
We review additional related work.

\paragraph{Spanning Tree Congestion.} 
We begin by describing the existing work on STC in more detail. There is a large body of work on the combinatorics and extremal bounds of STC for various graph classes including complete and complete multipartite graphs, grids, hypercubes, $k$-outerplanar graphs, random graphs, and much more~\cite{ostrovskii2004minimal,kozawa2009spanning,law2009spanning,ostrovskii2011minimum,kozawa2011spanning,law2011spanning,bodlaender2011spanning,law2013spanning}. A global upper bound was obtained by \cite{chandran2018sqrtmn}, who gave an exponential-time (resp. poly-time) algorithm constructing a spanning tree with congestion $\sqrt{mn}$ (resp. $\Tilde{O}\left(\sqrt{mn}\right)$) and showed that this is tight by showing that there exist expander-like graphs whose spanning tree congestion is $\Theta\left(\sqrt{mn}\right)$.
On the complexity side, computing the minimum congestion spanning tree has been shown to be NP-hard even on planar graphs, graphs of maximum degree $3$, graphs with diameter $2$, and more, and deciding whether $\stc(G)\leq k$ is NP-complete already for every fixed $k\geq5$ which implies the previous best known hardness of approximation of $6/5$ \cite{otachi2010complexity,okamoto2011hardness,bodlaender2012parameterized,luu2025better,atalig2026two,otachi2026spanning}. Its parameterized complexity has also been studied: for example, it is not fixed-parameter tractable parameterized by treewidth \cite{lampis2024parameterized}. In a series of recent works on graphs with maximum degree $\Delta$, Kolman gave an $\Tilde{O}\left(\Delta n^{1-1/(\sqrt{\log n}+1)}\right)$ approximation \cite{kolman2024approximating} and later improved this to
$\Tilde{O}(\Delta)$ \cite{kolman2025deltaapx}.

% TODO: Identify the intended minimum-mean matroid cycle reference and its connection to spanning-tree distributions.

\paragraph{Other Related Problems.}
A related line of work studies distributions over spanning trees that minimize the maximum expected congestion on any edge. Such distributions with polylogarithmic expected congestion can be both computed in polynomial time~\cite{andersen2009interchanging} and used for oblivious routing \cite{racke2008optimal}. Note that, here, the congestion guarantee holds in expectation for each edge, so it does not guarantee a single spanning tree of low congestion.

The Euclidean minimum dilation tree problem is a similar minmax spanning tree problem that asks for a spanning tree of a set of points in the Euclidean plane that (roughly) minimizes the maximum stretch of a pair of points compared to their Euclidean distance. This problem similarly withstood attempts to develop nontrivial approximations, and very recently was shown to admit a sublinear approximation \cite{eppstein1999spanning,de2026dilation}.
Another class of problems of similar flavor to STC consists of the cut-constrained spanning tree problems. Whereas low congestion spanning trees try to guarantee many tree edges across each cut compared to the original graph, cut-constrained spanning trees try to ensure that each cut among a set of specified cuts has a bounded number of spanning tree edges crossing it. These problems have been studied extensively for various families of cuts \cite{olver2018chain,klein2023thin,klein2026thin}.

\paragraph{Concurrent work.} Concurrent and independent work of \cite{atalig2026sublinearapproximationsspanningtree} gave $\Tilde{O}\left(n^{8/9}\right)$ and $\Tilde{O}\left(n^{2/3}\right)$ approximations for spanning tree congestion, and a $2-\epsilon$ hardness of approximation assuming $\textsf{P}\neq\textsf{NP}$. Their algorithms use a similar idea of using confluent flows on components with a single high-degree vertex and the degree-based approximation algorithm for low-degree components. Their $\Tilde{O}\left(n^{8/9}\right)$ approximation relies on an algorithm for connected \textit{minmax} multiway cut (a problem introduced in \cite{tiwary2026min}), and their $\Tilde{O}\left(n^{2/3}\right)$ approximation relies on a certain laminar decomposition.

\section{Definitions and Notation}\label{subsec:notation}
We begin with the (mostly standard) definitions and notation that we use.

\subsection{Graph Definitions and Notation}
Fix a graph $G=(V,E)$, and let $n:=|V|$ and $m:=|E|$. If $G$ is the input to a problem we will assume it is connected. Also, later, we omit subscripts when the graph is clear from context.

\paragraph{Basic notation.} 
Given $S\subseteq V$, we let $G[S]$ be the induced subgraph on $S$ in $G$. For a vertex $v$ we let $N(v):=\{u:\{u,v\}\in E\}$ denote $v$'s neighborhood. We let $\deg_G(v)$ denote the degree of vertex $v$ in $G$. The volume of a subset $S\subseteq V$ is $\vol_G(S)=\sum_{v\in S}\deg_G(v)$, and we write $\vol(G):=\vol_G(V)$. We let $\Delta_G(S) := \max_{u \in S} \deg_G(u)$ be the maximum degree of a vertex in $S$. Note that this may not be the same as the maximum degree of the graph $G[S]$. Likewise, we let $\Delta_G := \Delta_G(V)$.

\paragraph{Boundaries.} Given $S\subseteq V$, we let $\delta_G(S):=\{\{u,v\}\in E(G):u\in S,v\not\in S\}$ denote the edge boundary of $S$ in $G$. For a vertex $v$, we write $\delta_G(v):=\delta_G(\{v\})$.
We define the \textit{edge expansion} of $G$ by
\[
    \beta(G):=\min_{\substack{\varnothing\neq S\subseteq V\\ |S|\leq |V|/2}} \frac{|\delta_G(S)|}{|S|}
\]
or as $0$ if $|V| = 1$.

\paragraph{Cuts.} Given disjoint sets $S,T\subseteq V$, an \textit{$S$-$T$ vertex cut} is a set $C\subseteq V$ such that every path from $S$ to $T$ contains a vertex of $C$. We allow $C$ to intersect $S$ or $T$. An \textit{$S$-$T$ edge cut} is a set $F\subseteq E$ such that every path from $S$ to $T$ contains an edge of $F$. We also refer to a vertex set $A\subseteq V$ with $S\subseteq A$ and $A\cap T=\varnothing$ as an $S$-$T$ edge cut, identifying it with its edge boundary $\delta_G(A)$.

\paragraph{Flows.} We will use the following conventions for graph flows. Given sources and sinks $S, T \subseteq V$, an $S$-$T$ flow is a function $f : \mcP(S, T) \to \mathbb{Z}_{\geq 0}$ where $\mcP(S, T)$ is the set of all paths whose first vertex is in $S$, whose last vertex is in $T$ and which contain exactly one vertex of $T$. The \textit{value of $f$} is the total amount of flow sent from $S$ to $T$, namely $\sum_{P\in\mcP(S,T)} f(P)$. The \textit{congestion of node $u$ in flow $f$} is the flow going through $u$, i.e.\ $\sum_{P \ni u} f(P)$. Node $s \in S$ \textit{sends} $x$ flow in $f$ if the sum of flow over paths starting at $s$ is $x$, namely $\sum_{P \in \mcP(\{s\}, T)} f(P) = x$. A flow is \textit{confluent} if every $v\in V-T$ sends all of the flow at $v$ along a single edge, namely, for each $v\in V-T$, there is at most one $u\in N(v)$ such that $u$ immediately follows $v$ on a path $P\in\mcP(S,T)$ with $f(P)>0$. 
\subsection{Tree Definitions and Notation}
Next, fix a tree $T = (V,E)$.

\paragraph{Tree Cuts.} Given a vertex $v \in V$ (resp. an edge $e\in E$), we let $T^{-v}$ (resp. $T^{-e}$) be the graph resulting from deleting $v$ (resp. $e$) from $T$. Similarly, we let   $\left\{T^{-v}_i\right\}_i$ (resp. $\left\{T^{-e}_i\right\}_i$) denote the connected components of $T^{-v}$ (resp.\ $T^{-e}$).

\paragraph{Centroids.} Lastly, we make use of the well-known notion of a tree centroid.
\begin{definition}[Centroid]\label{defn:centroid}
    A centroid of a tree $T = (V,E)$ with respect to vertex set $S\subseteq V$ is a vertex $v\in V$ such that $\left|V(T^{-v}_i)\cap S\right|\leq |S|/2$ for every $i$.
\end{definition}
\noindent It is well-known that every tree has a centroid for any $S$; see, for example, \cite{berendsohn2023centroid}.

\section{Prior Work We Use}\label{sec:prelim}
We next review several known algorithmic results as well as simple corollaries that we will use. %We provide our notation in \Cref{subsec:notation}, reduction to a sublinear-on-$m$ approximation in \Cref{subsec:ntom}, subroutines for the decomposition in \Cref{subsec:mwc,subsec:cutmatch}, subroutine for high-degree components in \Cref{subsec:confluent}, and subroutine for low-degree components in \Cref{subsec:degapx}.

\subsection{Reducing a Sublinear in $n$ Approximation to a Sublinear in $m$ Approximation}\label{subsec:ntom}
First, we observe that prior work implies that an $m^{\epsilon}$ approximation for spanning tree congestion for any constant $\epsilon>0$ implies an $n^{\delta}$ approximation for some constant $\delta>0$ depending on $\epsilon$, as below. 

\begin{lemma}\label{lemma:bestofboth}
    If there is a poly-time $\Tilde{O}\left(m^{\epsilon}\right)$ approximation for spanning tree congestion for $\epsilon>0$, then there is a poly-time $\Tilde{O}\left(n^{\delta}\right)$ approximation where $\delta=\frac{3\epsilon}{2\epsilon+1}$. 
\end{lemma}
\noindent This follows essentially immediately from the below absolute lower and upper bounds on any graph's spanning tree congestion.
\begin{lemma}[STC Absolute Lower Bound \cite{otachi2011designing}]\label{lemma:globallowerbound}
    For any graph $G$ we have $\stc(G)\geq m/n$.
\end{lemma}
\begin{theorem}[STC Absolute Upper Bound \cite{chandran2018sqrtmn}]\label{thm:hardbound}
    There is a poly-time algorithm that, given a graph $G$,  outputs a spanning tree of $G$ with congestion at most $\Tilde{O}\left(\sqrt{mn}\right)$. 
\end{theorem}
% \begin{corollary}[Upper bound as an approximation]\label{coro:hardbound}
%     Given a graph $G$, there is a polynomial time algorithm that outputs a spanning tree of $G$ with congestion at most $\Tilde{O}\left(n^{3/2}/\sqrt{m}\right)\cdot\stc(G)$. 
% \end{corollary}

% \begin{proof}
%     By \Cref{lemma:globallowerbound}, we have $m/n\leq \stc(G)$ so the tree returned by \Cref{thm:hardbound} gives a $\sqrt{mn}/\stc(G)\leq \frac{\sqrt{mn}}{m/n}=\frac{n^{3/2}}{\sqrt{m}}$ approximation.
% \end{proof}

\noindent Using the algorithm of \Cref{thm:hardbound} when the graph is dense and the $m^\eps$-approximation when the graph is sparse gives our $n^\delta$ approximation.
\begin{proof}[Proof of \Cref{lemma:bestofboth}]
    First, observe that the above absolute bounds give a poly-time $\Tilde{O}\left(n^{3/2}/\sqrt{m}\right)$-approximation. In particular, by \Cref{lemma:globallowerbound}, we have $m/n\leq \stc(G)$ so the tree returned by \Cref{thm:hardbound} gives an approximation of $\Tilde{O}\left(\sqrt{mn}/\stc(G)\right)\leq \Tilde{O}\left(\frac{\sqrt{mn}}{m/n}\right)=\Tilde{O}\left(\frac{n^{3/2}}{\sqrt{m}}\right)$.

    Suppose we output whichever of the two trees returned by the $m^{\epsilon}$-approximation and the $\Tilde{O}\left(n^{3/2}/\sqrt{m}\right)$-approximation has smaller congestion. Letting $m=n^c$ we have that the resulting algorithm has an approximation bound of 
    \begin{align*}
        \min\left\{\Tilde{O}\left(n^{\epsilon c}\right),\Tilde{O}\left(n^{3/2-c/2}\right)\right\}.
    \end{align*}
    This is maximized when the exponents are equal at $c=\frac{3}{2\epsilon+1}$, for which both terms are $\Tilde{O}\left(n^{\frac{3\epsilon}{2\epsilon+1}}\right)$. 
\end{proof}
\noindent
An immediate consequence of this is our $\tilde{O}(n^{6/7})$-approximation, given a $\tilde{O}(m^{2/3})$-approximation.
\thmsublinear*
\begin{proof}
    This is immediate from \Cref{thm:sparse,lemma:bestofboth}.
\end{proof}
\noindent Thus, we focus on proving \Cref{thm:sparse}.

\subsection{A $2$-Approximation for Min Multiway Cut}\label{subsec:mwc}
Next, we state a textbook $2$-approximation algorithm for multiway cut. A multiway cut is as below. 
\begin{definition}[Multiway Cut]
    Given a graph $G=(V,E)$ and terminals $S\subseteq V$, a multiway cut of $S$ is a set of edges $F\subseteq E$ such that each connected component of $G-F$ contains at most one vertex of $S$. 
\end{definition}
\noindent The $2$-approximation for multiway cut is as below---for completeness, we give a proof.
\begin{lemma}[$2$-Approximation for Min Multiway Cut \cite{dahlhaus1994complexity,vazirani2001approximation}]\label{lemma:prelimmwc}
    Given a graph $G=(V,E)$ and terminals $S\subseteq V$ with $F^*$ the minimum-size multiway cut of $S$, there is a poly-time algorithm that outputs a multiway cut $F$ of $S$ satisfying $|F|\leq 2\left|F^*\right|$. 
\end{lemma}
\begin{proof}
    For each $v \in S$, let $F_v\subseteq E$ be a minimum cut between $v$ and $S-v$. Return $F=\bigcup_{v\in S}F_v$. This requires solving min cut at most $n$ times and so is poly-time. Likewise, each component of $G-F$ has at most one vertex of $S$ by construction.  
    To see the approximation bound, for each $v\in S$, let $C_v$ be the vertex set of the component of $G-F^*$ containing $v$. Since $\delta(C_v)$ separates $v$ from $S-v$, minimality of $F_v$ gives $|F_v|\leq|\delta(C_v)|$. Each $\delta(C_v)$ is contained in $F^*$, and each edge of $F^*$ belongs to at most two such boundaries because the $C_v$'s are disjoint. Thus
    \[
        |F|\leq\sum_{v\in S}|F_v|\leq\sum_{v\in S}|\delta(C_v)|\leq 2|F^*|. \qedhere
    \]
\end{proof}

\subsection{Global Degree Approximations for Subgraphs}\label{subsec:degapx}
\cite{kolman2025deltaapx} gave an $\tilde{O}(\Delta_G)$-approximation algorithm for spanning tree congestion. Here, we observe that, for any vertex set $L$ with $G[L]$ connected, the same algorithm can be run on $G[L]$ to get a spanning tree with congestion competitive with that of the minimum congestion spanning tree \emph{in $G$} (instead of just $G[L]$). More specifically, the algorithm loses in approximation the larger of $\Delta(L)$ (the maximum degree of a vertex of $L$ in $G$) and the number of edges crossing $L$. The formal statement is below.
\begin{theorem}[Global Degree Approximations for Subgraphs]\label{thm:extralocal}
    Given a graph $G = (V,E)$ and a vertex set $L\subseteq V$ such that $G[L]$ is connected, there is a poly-time algorithm that outputs a spanning tree of $G[L]$ with congestion at most 
    $\Tilde{O}\left(\max\left\{\Delta_G(L),|\delta_G(L)|\right\}\right)\cdot\stc(G)$. 
\end{theorem}
\noindent
\cite{kolman2025deltaapx} gave an $\tilde{O}(\Delta_G)$-approximation by reducing finding a low congestion spanning tree to arguing that every subgraph has bounded expansion, as below.\footnote{Their algorithm as stated in their paper is randomized because it uses \cite{arora2009expander}'s randomized algorithm for sparsest cut as a blackbox, but we can replace it with e.g.\ \cite{leighton1999multicommodity}'s deterministic algorithm in a blackbox fashion that only changes polylogarithmic factors in the approximation, which are negligible for our purposes.}
\begin{theorem}[Expansion Bounds Give STC Approximations \cite{kolman2025deltaapx,kolman2004crossing}]\label{lemma:kolmangeneral}
    Suppose $G$ is a graph such that for every nonempty subgraph $H$ of $G$,
    $\beta(H)\leq X/|V(H)|$. Then, there is a poly-time algorithm that outputs a spanning tree of $G$ with congestion at most $\Tilde{O}(X)$.
\end{theorem}
\noindent \cite{kolman2025deltaapx} applied this reduction by showing that every nonempty subgraph $H$ has expansion at most $\tilde{O}(\Delta \cdot \stc(G)/|V(H)|)$, thereby showing that the above is an $\tilde{O}(\Delta)$ approximation. 

In order to show \Cref{thm:extralocal}, we prove a similar fact. Namely, we bound the expansion of any subgraph of an induced graph $G[L]$ in terms of $\stc(G)$ (instead of $\stc(G[L])$ as applying \cite{kolman2025deltaapx}'s fact would).
\begin{lemma}[Expansion Bounds for Subgraphs in Terms of Global Degree]\label{lemma:extralocal}
    Given a graph $G = (V,E)$, any $L\subseteq V$ and any nonempty subgraph $H$ of $G[L]$, we have
    \[
        \beta(H)\leq \frac{3\max\{\Delta(L),|\delta(L)|\}\cdot\stc(G)}{|V(H)|}.
    \]
\end{lemma}
\begin{proof}
    If $|V(H)|=1$, the result is immediate since $\beta(H)=0$. Hence assume $|V(H)|\geq2$.
    It suffices to show there exists a cut $C \subseteq V(H)$ such that
    \[
        |V(H)|/3\leq |C|\leq |V(H)|/2
        \qquad\text{and}\qquad
        |\delta_H(C)|\leq \max\{\Delta(L),|\delta(L)|\}\cdot\stc(G).
    \]
    Indeed, $\beta(H)\leq |\delta_H(C)|/|C|$ by definition so the above conditions suffice. 
    
    Let $S:=V(H)$ and let $T^*$ be a minimum congestion spanning tree in $G$. Let $c$ be a centroid of $T^*$ with respect to $S$. Also recall $T^{*-c}_i$ is one of the connected components of $T^*-c$.

    We construct $C$ from $T^*$ as follows. First, we partition $V(H)$ into two sets $A$ and $B$. In particular, if $c \in V(H)$ we place it in $B$ and then, for each $i$, we place $V(H) \cap V(T^{*-c}_i)$ in one of $A$ or $B$ so that $|S|/3\leq |A\cap S|\leq 2|S|/3$. Such a partition is possible since $|V(T^{*-c}_i) \cap S| \leq |S|/2$ for each $i$. Furthermore, it follows that $|B\cap S|\leq 2|S|/3$.  Then, let $C$ be the smaller of $A\cap S$ and $B\cap S$. Observe that $|C| \leq |V(H)|/2$ by construction.
    
    We now bound $|\delta_H(C)|$. Towards this, observe that at most $\max\left\{\Delta(L),|\delta(L)|\right\}$ components $T^{*-c}_i$ contain a vertex of $S$; call such components \textit{relevant}. To see this, we case on $c\in L$.
    \begin{itemize}
        \item Suppose $c\in L$, then $\left|\left\{T^{*-c}_i\right\}_i\right|\leq\deg(c)\leq \Delta(L)$.
        \item Suppose $c\not\in L$. Consider a relevant component $T^{*-c}_i$ and let $T_i$ be the union of $T^{*-c}_i$ with the edge connecting it to $c$. All $T_i$ are connected and edge-disjoint by construction. Furthermore, each $T_i$ contains a vertex in $L$ (namely something in $S$) and a vertex not in $L$ (namely $c$). It follows that the number of $T_i$'s, and therefore the number of relevant components, is at most $|\delta(L)|$.
    \end{itemize}
    Additionally, observe that 
    \begin{equation}\label{equ:refinedlemmafirstineq}
        \left|\delta_G\left(V(T^{*-c}_i)\right)\right|\leq \stc(G)
    \end{equation} for any $i$ since every edge $\{u,v\}\in \delta_G\left(V(T^{*-c}_i)\right)$ uses the edge connecting $c$ to $T^{*-c}_i$ in the unique $u$ to $v$ path in $T^*$. 

    It follows that we can crudely bound $|\delta_H(C)|$ by summing over all relevant components   as
    $$ |\delta_H(C)| \leq \sum_{\text{$i$}} \left|\delta_G\left(V(T^{*-c}_i)\right)\right|\leq \sum_{\text{$i$}} \stc(G)\leq \max\left\{\Delta(L),|\delta(L)|\right\}\cdot \stc(G)$$
    where the second inequality is due to \eqref{equ:refinedlemmafirstineq} and the third inequality is because there are at most $\max\left\{\Delta(L),|\delta(L)|\right\}$ relevant components. 
\end{proof}
\noindent \Cref{thm:extralocal} is immediate from combining \Cref{lemma:kolmangeneral} and \Cref{lemma:extralocal}.

\subsection{Rounding Degree-to-Root Flows to Spanning Trees with Confluent Flows}\label{subsec:confluent}
Lastly, we will make use of prior work of \cite{chandran2018sqrtmn} which observed that certain low congestion flows can be rounded into low congestion spanning trees by using confluent flow.

Specifically, here, relevant flows are degree-to-root flows, where each node sends its degree in flow to some specified root.
\begin{definition}[Degree-to-Root Flow]
    Given a graph $G$, a subgraph $H$ of $G$, and a root $t\in V(H)$, a degree-to-root flow in $H$ with respect to $G$ is a flow $f$ in $H$ in which each vertex $v\in V(H)\setminus\{t\}$ sends exactly $\deg_G(v)$ flow to $t$. 
\end{definition}
\noindent We define the node congestion of a degree-to-root flow as $\max_{u\in V(H)\setminus\{t\}} \sum_{P\ni u} f(P)$. Note that we exclude the root when measuring congestion. Also, note that the flow is in the subgraph $H$.

The relevant confluent flow rounding fact is as below. 
\begin{theorem}[Confluent Flow Rounding \cite{chen2007confluent}]\label{thm:confluent}
    There is a poly-time algorithm that, given a graph $G=(V,E)$, sources $S\subseteq V$, a nonempty sink set $T\subseteq V$, and an $S$-$T$ flow $f$ with maximum node congestion at most $c$, outputs a confluent $S$-$T$ flow $f'$ such that each $s\in S$ sends the same flow as in $f$, and $f'$ has maximum node congestion at most $\Tilde{O}(c)$.
\end{theorem}
\noindent 
We now restate the observation of \cite{chandran2018sqrtmn} that, when applied to degree-to-root flows, the above confluent flow rounding gives bounded congestion spanning trees. This is implicit in \cite{chandran2018sqrtmn} but not directly stated; as such we give a proof for completeness.
\begin{lemma}[Rounding Degree-to-Root Flows to Trees \cite{chandran2018sqrtmn}]\label{lemma:feasibleflow}
        There is a poly-time algorithm that, given a flow to a root in a graph $G=(V,E)$ in which each non-root vertex $v$ sends at least $\deg_G(v)$ flow and the node congestion outside the root is at most $c$, outputs a spanning tree of $G$ with congestion $\Tilde{O}(c)$.
\end{lemma}
\begin{proof}
    We will construct a tree from a confluent rounding of the given flow with root $t$.
    
    First, if $V=\{t\}$, we return the one-vertex tree. Otherwise, let $Q=N(t)$ and truncate each flow path starting outside $t$ at its first vertex in $Q$. This gives a $(V\setminus\{t\})$-$Q$ flow in $G-t$ with maximum node congestion at most $c$, including at its sinks. Each vertex sends the same flow by construction. Apply \Cref{thm:confluent} to this flow, and then extend each resulting path ending at $q\in Q$ by the edge $\{q,t\}$. The resulting flow $f$ is a confluent flow to $t$ in $G$ with node congestion $\Tilde{O}(c)$ outside $t$, and every $q\in Q$ sends its flow directly to $t$.

    We next construct a spanning tree from the confluent flow $f$. For each $q\in Q$, let $V_q\subseteq V\setminus(\{t\}\cup Q)$ be the subset of nodes whose flow enters $t$ along $\{q,t\}$. Since every non-root vertex sends positive flow and $f$ is confluent, the sets $V_q\cup\{q\}$ form a partition of $V\setminus\{t\}$. Moreover, every vertex other than $q$ on the flow path from $v\in V_q$ to $q$ is also in $V_q$, so $G[V_q\cup\{q\}]$ is connected. Since each node sends at least its degree, the congestion at $q$ gives
    $\vol_G\left(V_q\cup\{q\}\right)\leq\sum_{P\ni q}f(P)\leq\Tilde{O}(c)$.
    We choose an arbitrary spanning tree $T_q$ of each $G\left[V_q\cup\{q\}\right]$, and return the tree $T$ obtained by joining each $T_q$ to $t$ by $\{q,t\}$. Clearly, $T$ is a spanning tree of $G$.

    We lastly prove the spanning tree congestion claim. For any $e\in E(T)$, let the side of $T^{-e}$ not containing $t$ be $S$. We have by construction that $S\subseteq V_q\cup\{q\}$ for some $q\in Q$. Thus $\vol_G(S)\leq\vol_G\left(V_q\cup\{q\}\right)\leq\Tilde{O}(c)$. The congestion of $e$ is $|\delta_G(S)|\leq\vol_G(S)$, proving that the congestion of $T$ is also at most $\Tilde{O}(c)$. The algorithm is trivially poly-time.
\end{proof}

\section{Degree-to-Root Cutmatches}\label{subsec:cutmatch}
In this section, we observe that the parts of a graph that do not admit a low congestion degree-to-root flow can always be separated from the root with a sparse cut. 

\begin{figure}[H]
    \centering
    \begin{subfigure}[t]{0.24\textwidth}
        \centering
        \includegraphics[width=\linewidth]{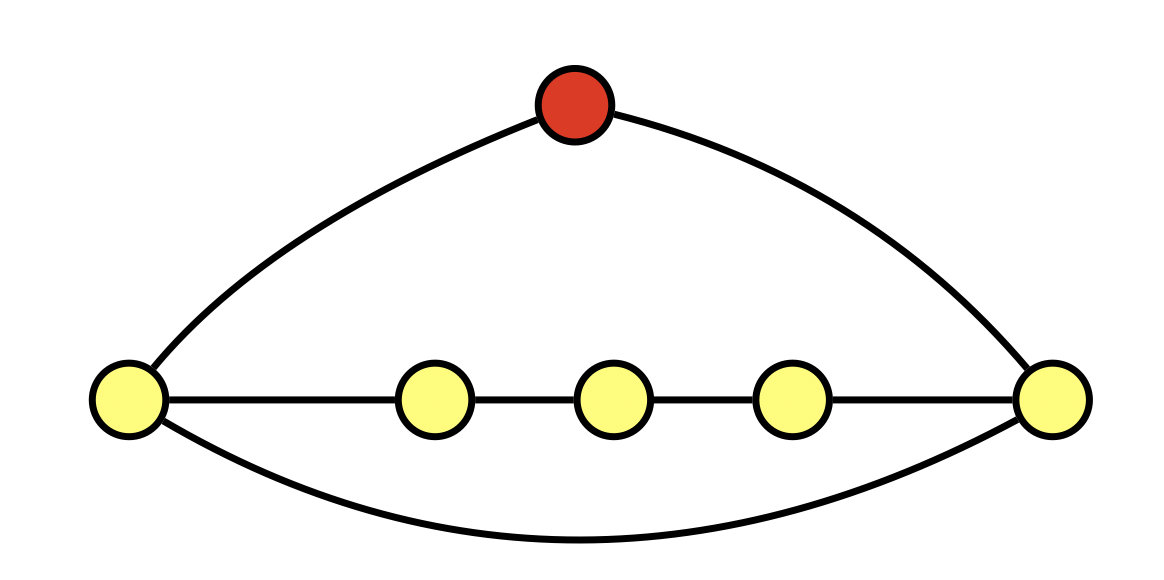}
        \caption{Graph $G$.}
                \label{subfig:flow1}
    \end{subfigure}\hfill%
    \begin{subfigure}[t]{0.24\textwidth}
        \centering
        \includegraphics[width=\linewidth]{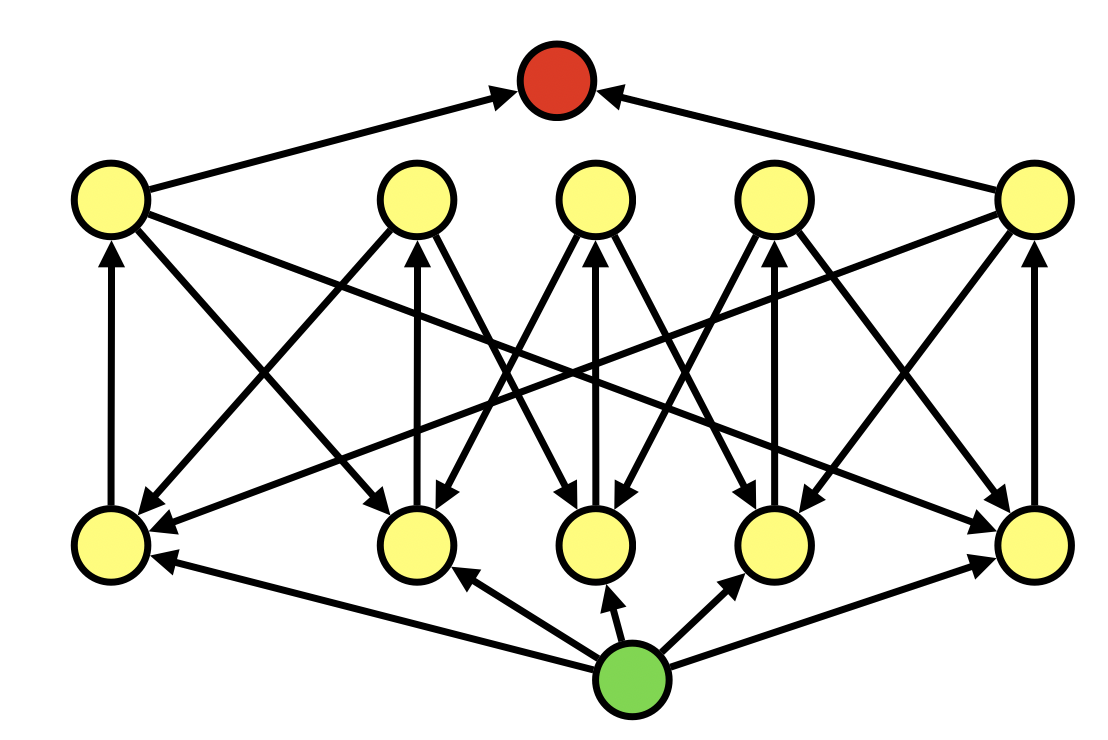}
        \caption{Graph $G'$.}
                \label{subfig:flow2}
    \end{subfigure}\hfill%
    \begin{subfigure}[t]{0.24\textwidth}
        \centering
        \includegraphics[width=\linewidth]{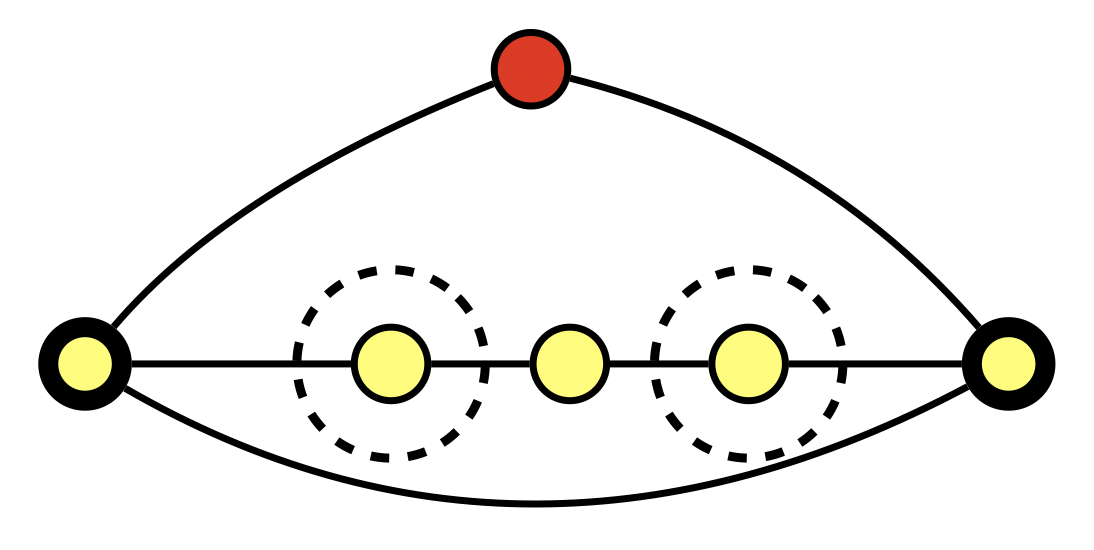}
        \caption{A cutmatch in $G$.}
                \label{subfig:flow3}
    \end{subfigure}\hfill%
    \begin{subfigure}[t]{0.24\textwidth}
        \centering
        \includegraphics[width=\linewidth]{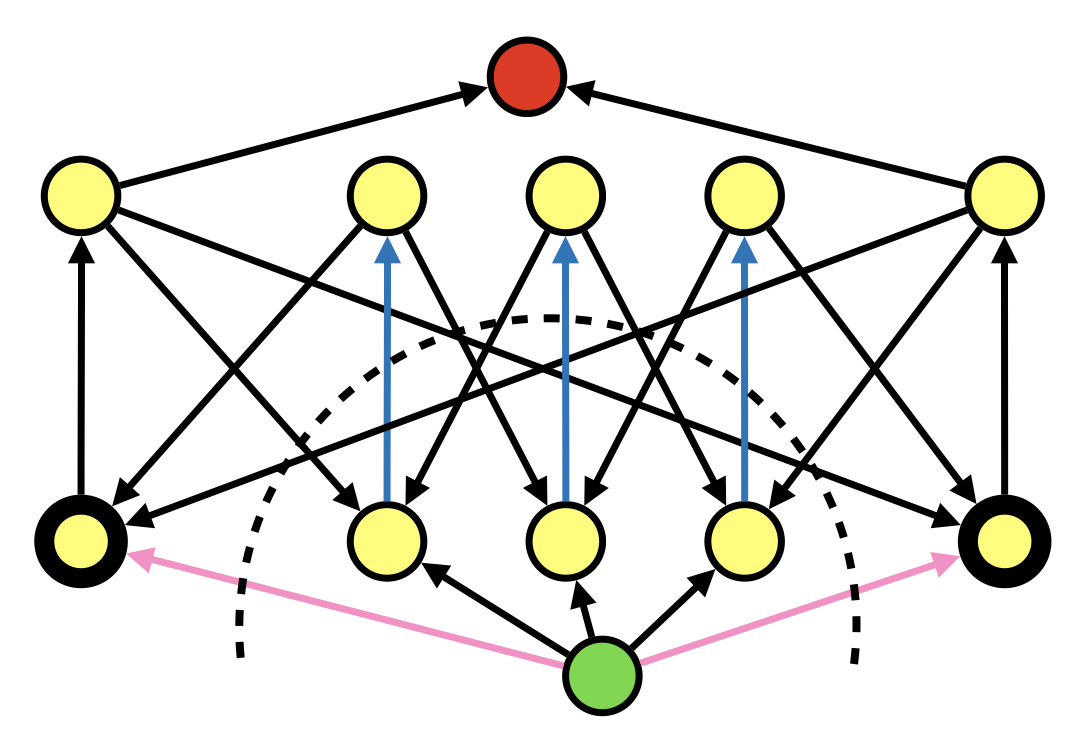}
        \caption{An $s-t$ mincut in $G'$.}
                \label{subfig:flow4}
    \end{subfigure}
    \caption{The construction of $G'$. \Cref{subfig:flow1}: a graph $G$ where $t$ is the red vertex. \Cref{subfig:flow2}: the corresponding $G'$, where $s$ is the green vertex and $t$ is the red vertex. The arcs from $s$ have capacity the degree of the other endpoint in $G$, the upwards copy arcs have capacity $c$, the downward arcs have capacity $\infty$, and the arcs to $t$ have capacity $\infty$. \Cref{subfig:flow3,subfig:flow4}: a minimum $s$-$t$ cut in $G'$ consists of upwards copy arcs in blue (whose in copies correspond to nodes in $U$) and arcs from $s$ to bolded in copies (corresponding to nodes in $M$, in bold). The vertex cut $C$ is marked with dotted circles.}
    \label{fig:flownetwork}
\end{figure}

More formally, we show that a (poly-time computable) cutmatch always exists. A cutmatch for degree-to-root flows partitions vertices into ``matched'' vertices $M$ and ``unmatched'' vertices $U$ where the matched vertices have a low congestion degree-to-root flow and the unmatched vertices are separated from the root by a sparse cut.
\begin{definition}[Degree-to-Root Cutmatch] \label{defn:cutmatch}
A congestion $c$ degree-to-root cutmatch of graph $G = (V,E)$ and sink $t\in V$ is a partition $V-t=M\sqcup U$ and a vertex cut $C\subseteq U $ such that
\begin{itemize}
    \item \textbf{Flow for Matched.} $G\left[M\cup\{t\}\right]$ has a degree-to-root flow w.r.t.\ $G$ with root $t$ and node congestion at most $c$; and
    \item \textbf{Sparse Cut for Unmatched.} $C$ is a $U$-$\{t\}$ cut with $|C|\leq \frac{1}{c} \cdot \vol\left(U\right)$.
\end{itemize}
\end{definition}\noindent 
The following both demonstrates the existence of degree-to-root cutmatches and gives algorithms to efficiently compute them. It follows by max-flow-min-cut applied to a carefully constructed auxiliary graph.
\begin{lemma}[Cutmatch]\label{lemma:cutmatch}
    There is a poly-time algorithm that, given a graph $G=(V,E)$, sink $t\in V$ and $c \in \mathbb{Z}_{\geq 1}$, outputs a degree-to-root cutmatch with congestion $c$. 
\end{lemma}
\begin{proof}
    % If there is already a degree-to-root flow from $V$ to $t$ with node congestion $c$ then we set $M=V-t,U=\emptyset,C=\emptyset$ and we are done. Otherwise, 
    We compute our cutmatch from the mincut of an auxiliary graph of $G$ obtained by splitting each vertex into in and out copies. For this proof, given $S\subseteq V$ we let $\partial^- (S):=\left\{v\in S:\exists\text{ } u\in N(v) \cap (V\setminus S) \right\}$ denote the \textit{inner vertex boundary} of $S$ in $G$. Also, for a directed graph $H$ and $S\subseteq V(H)$, we define $\delta_H(S):=\{(u,v)\in E(H):u\in S,v\notin S\}$ to be the set of arcs leaving $S$.

    Specifically, our auxiliary directed graph $G'=(V',E')$ with capacities on arcs is as follows. $V'$ consists of a dummy source $s$, a sink $t$, and two node copies $v^+,v^-$ for every $v\in V-t$. $E'$'s arcs include $(s,v^-)$ with capacity $\deg(v)$ for every $v\in V-t$, $(v^-,v^+)$ with capacity $c$ for every $v\in V-t$, $(u^+,v^-),(v^+,u^-)$ with capacity $\infty$ for every $\{u,v\}\in E-\delta(t)$, and $(v^+,t)$ with capacity $\infty$ for every $v$ adjacent to $t$ in $G$. See also \Cref{fig:flownetwork}. Hence a path in $G'$ of the form $s,u^-,u^+,v^-,v^+,\dots,t$ corresponds to a path in $G$ of the form $u,v,\dots,t$. Observe that an $s$-$t$ flow in $G'$ that sends $\vol(V-t)$ flow from $s$ to $t$ and in which no arc carries more flow than its capacity corresponds to a degree-to-root flow in $G$ with node congestion $c$.

    Our cutmatch is computed from a min $s$-$t$ edge cut $S'$ containing $s$ of $G'$ as follows. $\delta_{G'}(S')$ contains two types of arcs: source arcs of the form $(s,v^-)$ and copy arcs of the form $(v^-,v^+)$ (and no other arcs because all other arcs have infinite capacity). Let $B:=\{v\in V\setminus\{t\}:v^-\in S'\}$ be all vertices whose in copy is in $S'$. 
    Note that every $v\in V\setminus(B\cup\{t\})$ corresponds to a source arc $(s,v^-)\in\delta_{G'}(S')$, every $v\in \partial^-(B)$ corresponds to a copy arc $(v^-,v^+)\in\delta_{G'}(S')$, and no other copy arc is cut by $S'$.
    
    We define the matched vertices, unmatched vertices and vertex cut of our cutmatch respectively as  
    $$M:=V-t-B \qquad  \text{and} \qquad U:=B \qquad \text{and} \qquad C:=\partial^-(B).$$
    We argue that $C$ is a sparse cut for the unmatched vertices $U$. $C$ is a vertex cut separating $B$ from $t$, so it is a $U$-$\{t\}$ vertex cut. To bound its size, observe that by max-flow-min-cut duality, we have that the total edge capacity of $\delta_{G'}(S')$ and the maximum edge-capacitated flow in $G'$ are both $\vol(V-t-B)+c|C|$. On the other hand, observe that the maximum flow value in $G'$ is at most $\vol(V-t)$, the total capacity of the arcs leaving $s$. Thus, we have
    $$\vol(V-t-B) + c|C| \leq \vol(V-t) = \vol(V-t-B) + \vol(B)$$
    and subtracting $\vol(V-t-B)$ on the left and right sides gives $c|C|\leq \vol(B)$ as desired.

    We lastly prove that there is a degree-to-root flow in $G[M\cup\{t\}]$ with root $t$ and congestion $c$. Let $f$ be the maxflow in $G'$ computed to get $S'$. 
    A unit of flow either originates from $V-t-B$, for which there are at most $\vol(V-t-B)$ such units of flow, or originates in $B$, in which case it must pass through a vertex of $C=\partial^-(B)$ by definition. Since the copy arcs $(v^-,v^+)$ for $v\in C$ have total capacity $c|C|$, there are at most $c|C|$ such units of flow.
    Furthermore $f$ sends exactly $\vol(V-t-B)+c|C|$ units of flow to $t$ by max-flow-min-cut duality. Then $V-t-B$ sends exactly $\vol(V-t-B)$ units of flow and $B$ sends exactly $c|C|$ units. This implies that all the flow on $C$ is sent from $B$, so the $\vol(V-t-B)$ flow originating from $V-t-B$ cannot enter $B$ and hence stays in $V-t-B$ until reaching $t$, implying the flow for matched property. 
\end{proof}

% \section{Main Algorithm}
% We first describe our decomposition, and then combine it with the subroutines from the preliminaries to get the final algorithm.
\section{Flow-Or-Degree Decomposition}\label{sec:decomp}
Using the cutmatches of the preceding section, we now prove that, after deleting a small number of edges, every graph can be broken into connected components each of which either admits a low congestion degree-to-root flow or has low degree.
Our graph decomposition is as follows.
\begin{restatable}[Flow-Or-Degree Decomposition]{lemma}{lemmadecomp}\label{lemma:decomp}
    Given a graph $G=(V,E)$ with $m$ edges, there is a poly-time algorithm that computes a set $I\subseteq E$ of $O(m^{2/3})$ edges such that each connected component of $G-I$, with vertex set $C$, satisfies either
    \begin{itemize}
        \item \textbf{Low Degree.} $\Delta_G(C) \leq O(m^{2/3})$ or
        \item \textbf{Low Congestion.} $G[C]$ has a degree-to-root flow w.r.t.\ $G$ of node congestion $O(m^{2/3})\cdot\stc(G)$ for a root in $C$.
    \end{itemize}
\end{restatable}

% \rnote{replace decomposition with the decided name. i propose multiway cutmatch with congestion $c$ ($c$-multiway cutmatch and multiway $c$-cutmatch seem to emphasize $c$ at the wrong parts). one of $c,D$ must be specified because we test every $D$}
% \begin{definition}[Decomposition]\label{defn:decomp}
%     Given a graph $G=(V,E)$ and terminals $S\subseteq V$, a $c$-decomposition is a connected multiway cut $I\subseteq E$ of $G$ 
%     such that every connected component of $G-I$ containing a vertex in $S$ admits a degree-to-root flow with node congestion $c$.
% \end{definition}
\noindent Our decomposition uses multiway cuts and so we start with a simple observation that the size of a multiway cut from the preliminaries can be bounded by the spanning tree congestion. 
\begin{lemma}[Multiway Cut in Terms of STC]\label{lemma:existmwc67}\label{lemma:multiwaycut}
    There is a poly-time algorithm that, given a graph $G=(V,E)$ and $H\subseteq V$, finds a multiway cut $F\subseteq E$ of $H$ satisfying $|F|\leq O(|H|)\cdot\stc(G)$.
\end{lemma}
\begin{proof}
    Our algorithm simply takes the multiway cut $F$ of $H$ that is a $2$-approximation as given by \Cref{lemma:prelimmwc}.

    To see that $|F|\leq O(|H|)\cdot\stc(G)$, we argue that the min multiway cut has size at most $O(|H|)\cdot\stc(G)$. To do so, let $T^*$ be a minimum congestion spanning tree of $G$ and fix an arbitrary $h^*\in H$. For $h\in H\setminus\{h^*\}$, let $e_h$ be the first edge on the unique $h$ to $h^*$ path in $T^*$, and let $T^*_h$ be the component of $T^*-e_h$ containing $h$. The edges of $\delta_G(V(T^*_h))$ are exactly the edges of $G$ whose paths in $T^*$ use $e_h$, so $|\delta_G(V(T^*_h))|\leq\stc(G)$. Therefore, $\tilde{F}:=\bigcup_{h\in H\setminus\{h^*\}}\delta_G(V(T^*_h))$ has $$|\tilde{F}|\leq (|H|-1)\stc(G).$$ Each component of $G-\tilde{F}$ contains at most one vertex of $H$ by construction and so $\tilde{F}$ is a feasible multiway cut of $H$. Since $\tilde{F}$ is a feasible multiway cut of $H$, we have $$|F|\leq 2|\tilde{F}|\leq 2(|H|-1)\stc(G)$$ as needed.
\end{proof}
 \noindent
Our decomposition takes a multiway cut of all high degree nodes and then further breaks up the components with high degree nodes by taking a cutmatch. Ideally, we would take a cutmatch with congestion $c=m^{2/3}\cdot\stc(G)$ but since we may not know $\stc(G)$ we try increasing guesses $c=m^{2/3},m^{2/3}+1,\ldots$, using degree threshold $c/m^{1/3}$, and return the first successful guess. Our formal algorithm is given by \Cref{alg:decomp} where we suppress constants with $O$ for the sake of presentation.
\begin{algorithm}[H]
    \caption{Flow-Or-Degree Decomposition}\begin{algorithmic}[1]
        \label{alg:decomp}
    \Statex \textbf{Input}: Graph $G = (V,E)$.
    \Statex\textbf{Output}: Flow-or-degree decomposition $I\subseteq E$.
    \For{$c=m^{2/3},m^{2/3}+1,\dots,m$}
        \State Let $H:=\{v\in V(G):\deg_G(v)\geq c/m^{1/3}\}$ be all high degree nodes.
        \State Compute a multiway cut $F$ of $H$ by \Cref{lemma:multiwaycut}.
        \For{$h\in H$ with component vertex set $P_h$ in $G-F$}
            \State Let $C_h$ be the cut of the cutmatch from \Cref{lemma:cutmatch}
        on $(G-F)[P_h]$ with sink $h$ and congestion $c$.
        \EndFor
        \State Let $I:=F\cup \bigcup_h \delta\left(C_h\right)$.
        \State If $|I|\leq O(m^{2/3})$, \textbf{return} $I$.\label{line:decomp-return}
    \EndFor
    \State \textbf{Return} $I=\varnothing$.
    \end{algorithmic}
\end{algorithm}
\noindent
We conclude this section by proving that \Cref{alg:decomp} gives our decomposition.
\begin{proof}[Proof of \Cref{lemma:decomp}]
    We use \Cref{alg:decomp}. The algorithm is trivially poly-time by the guarantees of \Cref{lemma:multiwaycut} and \Cref{lemma:cutmatch}.

    First, observe that the $I$ that we return consists of $O(m^{2/3})$ edges by construction. Thus, it suffices to argue that each connected component of $G-I$ satisfies either the low degree or the low congestion guarantee.
    
    First, we claim that if $m^{2/3}\cdot\stc(G) > m$ then no matter what $I$ we return we trivially satisfy the low congestion guarantee on every connected component of $G-I$. In particular, for each component $C$, choose an arbitrary root $r\in C$ and a flow in $G[C]$ in which each vertex $v \in C-r$ sends $\deg_G(v)$ flow to $r$. This is a degree-to-root flow with respect to $G$, and every non-root vertex has node congestion at most the total flow sent which is $O(m) \leq O(m^{2/3}\cdot\stc(G))$.

    On the other hand, suppose $m^{2/3}\cdot\stc(G) \leq m$. We will argue that there is some $c\leq m^{2/3}\cdot\stc(G)$ which will cause us to return $I$ on line~\ref{line:decomp-return} of \Cref{alg:decomp}. Note that if this holds then $|I| \leq O(m^{2/3})$ by construction. Moreover, applying the guarantees of \Cref{lemma:multiwaycut} and \Cref{lemma:cutmatch}, we would have that each component $C$ of $G-I$ either contains no vertex of $H$ in which case $\Delta_G(C) \leq c/m^{1/3} \leq  O(m^{2/3})$ or $C$ consists of the matched vertices of one of our cutmatches along with some $h \in H$. We claim that, in the latter case, $G[C]$ has a degree-to-root flow with respect to $G$ with root $h$ of node congestion $O(m^{2/3})\cdot\stc(G)$. Indeed, in this case let $J_h:=(G-F)[P_h]$. The cutmatch on $J_h$ gives a degree-to-root flow in $J_h[C]$ with respect to $J_h$ of node congestion at most $c$. However, we need a degree-to-root flow in $G[C]$ with respect to $G$. We construct this by routing, for every $v\in C\setminus\{h\}$, an additional $\deg_G(v)-\deg_{J_h}(v)$ units of flow to $h$ along any path in $G[C]$. This guarantees that each $v \in C$ other than $h$ routes $\deg_G(v)$ flow to $h$, and the node congestion at every node other than $h$ is at most $c+2|F|\leq c+2|I|=O(m^{2/3})\cdot\stc(G)$, as required.
    
    Thus, it remains to argue that for some $c\leq m^{2/3}\cdot\stc(G)$, we return on line~\ref{line:decomp-return}. To argue this, we argue that if $c$ gets to $m^{2/3}\cdot\stc(G)$ then we return on line~\ref{line:decomp-return}. In particular, we will argue that in this case we must have $|I| \leq O(m^{2/3})$. To do so, observe that since $I:=F\cup \bigcup_h \delta\left(C_h\right)$, it suffices to bound the cardinality of our multiway cut, $|F|$, and the cardinality of the edges leaving all cutmatch cuts, $|\bigcup_h \delta\left(C_h\right)|$ both as $O(m^{2/3})$.
    
    We first bound $|F|$ as $O(m^{2/3})$. By the definition of our high degree nodes $H$, we have $|H|\leq O(m^{4/3}/c) = O(m^{2/3}/\stc(G))$. Hence, \Cref{lemma:multiwaycut} guarantees that the total number of edges cut by our multiway cut of high degree nodes $H$ is
    \begin{align}\label{eq:rtesds}
        |F|\leq O(|H|)\cdot\stc(G)
        \leq O\left(\frac{m^{4/3}\cdot\stc(G)}{c}\right)
        =O(m^{2/3}).
    \end{align}
    We next bound $|\bigcup_h \delta\left(C_h\right)|$ as $O(m^{2/3})$. To do so, we first bound the degree of each vertex of $C_h$ and then bound the cardinality of each $C_h$. In particular, observe that since $P_h$ is a connected component after taking a multiway cut of $H$, we know that every vertex $u \in P_h$ which is not $h$ has degree $\deg_G(u) \leq c/m^{1/3} = m^{1/3} \cdot \stc(G)$. Since $C_h$ is just a subset of vertices of $P_h$ not including $h$, it follows that for each $u \in C_h$ we have 
    \begin{align}\label{eq:dhgsas}
        \deg_G(u) \leq m^{1/3} \cdot \stc(G).
    \end{align}
    Next, letting $U_h$ be the unmatched set of the cutmatch computed for $h$, we have by \Cref{lemma:cutmatch} that $|C_h| \leq \frac{1}{c} \cdot \vol_{(G-F)[P_h]}(U_h)$. Using $\vol_{(G-F)[P_h]}(U_h) \leq \vol_G(U_h)$ we get
    \begin{align}\label{eq:ryte}
        |C_h|\leq\frac{1}{c}\cdot\vol_G(U_h) = O\left( \frac{\vol_G(U_h)}{m^{2/3} \cdot \stc(G)} \right)
    \end{align}
    Combining \Cref{eq:dhgsas} and \Cref{eq:ryte} and applying the fact that the $P_h$ are pairwise disjoint, we get
    \begin{align}\label{eq:terewq}
        \left|\bigcup_h \delta\left(C_h\right)\right| \nonumber
        &\leq O\left(\frac{1}{m^{1/3}} \right) \sum_{h \in H}\vol_G(U_h)\\
        & = O(m^{2/3}).
    \end{align}
    Thus, we know that if $c = m^{2/3}\cdot\stc(G)$ then by \Cref{eq:rtesds} and \Cref{eq:terewq} we have $|I| \leq |F| + \left|\bigcup_h \delta\left(C_h\right)\right| \leq O(m^{2/3})$ which is to say that our algorithm returns on line~\ref{line:decomp-return} as required.
\end{proof}

\section{A $\tilde{O}(m^{2/3})$-Approximation for Spanning Tree Congestion}\label{sec:alg}
We now use the flow-or-degree decomposition of the preceding section to give our $\tilde{O}(m^{2/3})$-approximation for spanning tree congestion.

\thmmain*
\noindent
Roughly, the algorithm on graph $G=(V,E)$ is as follows. First, compute a flow-or-degree decomposition $I$ as in \Cref{lemma:decomp}. For each low degree component of $G-I$, construct a tree using the global degree approximation algorithm of \Cref{thm:extralocal}. For each component of $G-I$ with a low congestion degree-to-root flow, round the flow to a low congestion spanning tree using \Cref{lemma:feasibleflow}. Then, arbitrarily complete the resulting forest to a spanning tree. The pseudocode is given in \Cref{alg:main} where we suppress constants with $O$ notation for readability.
\begin{algorithm}[H]
    \caption{Main Algorithm}\begin{algorithmic}[1]
        \label{alg:main}
    \Statex \textbf{Input}: Graph $G=(V,E)$.
    \Statex \textbf{Output}: Spanning tree $T$ of $G$ satisfying $\con_G(T)\leq \Tilde{O}\left(m^{2/3}\right)\cdot \stc(G)$.
    \State Compute a flow-or-degree decomposition $I$ of $G$ using \Cref{lemma:decomp}.
    \State Let $\mathcal P$ be the vertex sets of the connected components of $G-I$.
    \For{$C\in\mathcal L:=\{C\in\mathcal P:\Delta_G(C)\leq O(m^{2/3})\}$} \Comment{Low-degree components}
        \State Apply \Cref{thm:extralocal} to $G$ and $C$ to obtain a spanning tree $T_C$ of $G[C]$.
    \EndFor
    \For{$C\in\mathcal R:=\mathcal P\setminus\mathcal L$} \Comment{Low-congestion components}
        \State Let $f_C$ be a degree-to-root flow in $G[C]$ with respect to $G$ of node congestion $O(m^{2/3})\cdot\stc(G)$.
        \State Apply \Cref{lemma:feasibleflow} to $f_C$ to obtain a spanning tree $T_C$ of $G[C]$.
    \EndFor
    
    \State Let $A\subseteq I$ be the edges of any spanning tree of the graph obtained from $G$ by contracting each $C\in\mathcal P$.
    \State Output the tree $T$ with edge set $A\cup\bigcup_{C\in\mathcal P}E(T_C)$.
    \end{algorithmic}
\end{algorithm}

\noindent We now prove that \Cref{alg:main} is an $\Tilde{O}\left(m^{2/3}\right)$ approximation for spanning tree congestion.

\begin{proof}[Proof of \Cref{thm:sparse}]
    We use \Cref{alg:main}. The algorithm is trivially poly-time by the guarantees of \Cref{lemma:decomp}, \Cref{thm:extralocal}, and \Cref{lemma:feasibleflow}, since $|\mathcal P|\leq n$ and the flows $f_C$ can be obtained by running polynomially many max-flow computations. Likewise, since each $T_C$ spans $G[C]$ and $A$ spans the contracted graph, $T$ is a spanning tree of $G$.
    
    It remains to prove the approximation guarantee. In particular, we claim that the output tree $T$ has congestion at most $\Tilde{O}(m^{2/3})\cdot\stc(G)$. We bound the congestion of each $e\in E(T)$ by considering where $e$ came from.
    \begin{itemize}
        \item If $e\in A$, the only edges which congest $e$ are those of $I$ so $e$ has congestion at most $|I| = O(m^{2/3})$ by \Cref{lemma:decomp}.
        \item If $e\in E(T_C)$ for $C\in\mathcal R$ then the only edges that congest $e$ belong to $I$ or $E(G[C])$. The contribution of the former to the congestion of $e$ is $O(m^{2/3})$ by \Cref{lemma:decomp} and the contribution of the latter is $\Tilde{O}(m^{2/3})\cdot\stc(G)$ by the congestion of the degree-to-root flow as in \Cref{lemma:decomp} and the additional congestion incurred by the rounding of \Cref{lemma:feasibleflow}.
        \item If $e\in E(T_C)$ for $C\in\mathcal L$, then the only edges that congest $e$ belong to $I$ or $E(G[C])$. The contribution of the former to the congestion of $e$ is again $O(m^{2/3})$ by \Cref{lemma:decomp} and, by \Cref{thm:extralocal}, the contribution of the latter is at most $\Tilde{O}\left(\max\{\Delta_G(C) ,|I|\}\right)\cdot\stc(G) = \Tilde{O}(m^{2/3})\cdot\stc(G)$ since in this case $\Delta_G(C) \leq O(m^{2/3})$ and $|\delta_G(C)| \leq |I| \leq O(m^{2/3})$ by \Cref{lemma:decomp}. \qedhere
    \end{itemize}
\end{proof}

% \noindent Taking the better of \Cref{alg:main} and \Cref{thm:hardbound} gives our final algorithm.

\section{Hardness from the Small Set Expansion Hypothesis}\label{sec:hard}
In this section, we show that, assuming the Small Set Expansion Hypothesis, it is NP-hard to approximate spanning tree congestion within any constant factor.
\lowerbound*
\noindent
We won't use the Small Set Expansion Hypothesis directly, but rather the following known consequence.
\begin{theorem}[Corollary of Theorem 3.5 of \cite{raghavendra2012reductions}, see also \cite{austrin2012inapproximability}]\label{thm:sseh}
    Assuming the Small Set Expansion Hypothesis is true, for every sufficiently small $\epsilon>0$ it is NP-hard to distinguish whether a graph $H$ satisfies
    \begin{itemize}
        \item \textbf{YES (Balanced Partition)}: there is a partition $V_1 \sqcup V_2 \sqcup \ldots$ of $V(H)$ such that $\vol_H(V_i)\leq\epsilon\cdot\vol(H)$ for every $i$ and $\left|\bigcup_i\delta_H(V_i)\right|\leq\epsilon\cdot\vol(H)$ or
        \item \textbf{NO (Balanced Sets Expand)}: every $S\subseteq V(H)$ satisfying $\vol(H)/10\leq\vol_H(S)\leq 9 \vol(H)/10$ also satisfies $|\delta_H(S)|\geq \Omega(\sqrt{\epsilon}\cdot\vol(H))$.
    \end{itemize}
\end{theorem}
\noindent In other words, YES means there exists a volume-balanced partition with few cut edges, and NO means every volume-balanced subset is expanding. 

At a high level, we take $H$ and construct a constant-degree graph $G$ that ``preserves'' the YES/NO answer of $H$, and then show that the spanning tree congestion of $G$ distinguishes between YES/NO of $G$. 
To do so, we construct a constant-degree graph $G$ by replacing every vertex of $H$ with a constant-degree expander, so $H$'s ``skeleton'' is preserved. We require constant degree in order to connect the spanning tree congestion of $G$ to the YES/NO cases, and we require constant-degree expanders (as opposed to any constant-degree graph) in order for $G$ to preserve the YES/NO answer of $H$.

We distinguish the YES and NO cases by proving upper and lower bounds on $\stc(G)$. In the YES case, we refine the balanced partition into connected parts and join arbitrary spanning trees of these parts to upper bound $\stc(G)$. Specifically, since $G$ has constant degree, each part has small volume; together with the small number of crossing edges, this gives congestion $O(\epsilon\cdot|V(G)|)$. In the NO case, we use a centroid argument to lower bound $\stc(G)$. Specifically, for any spanning tree, expansion forces $\Omega(\sqrt{\epsilon}\cdot|V(G)|)$ edges across a balanced cut. Then, we can form such a cut by grouping subtrees incident to the centroid and since the centroid has constant degree, this gives spanning tree congestion $\Omega(\sqrt{\epsilon}\cdot|V(G)|)$.

\begin{figure}[t]
    \centering
    \begin{subfigure}[t]{0.5\textwidth}
        \centering
        \includegraphics[width=.48\linewidth]{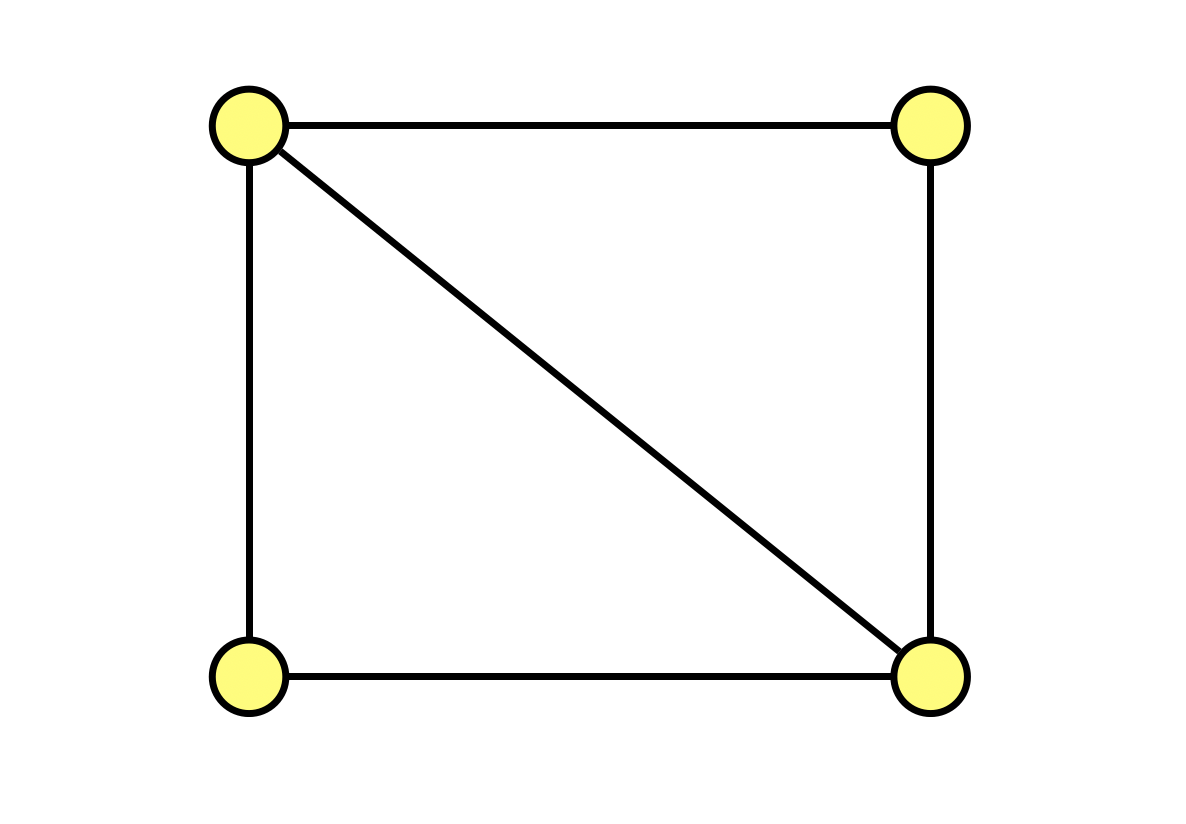}
        \caption{Graph $H$.}
                \label{subfig:sseh1}
    \end{subfigure}%
    ~
    \begin{subfigure}[t]{0.5\textwidth}
        \centering
        \includegraphics[width=.5\linewidth]{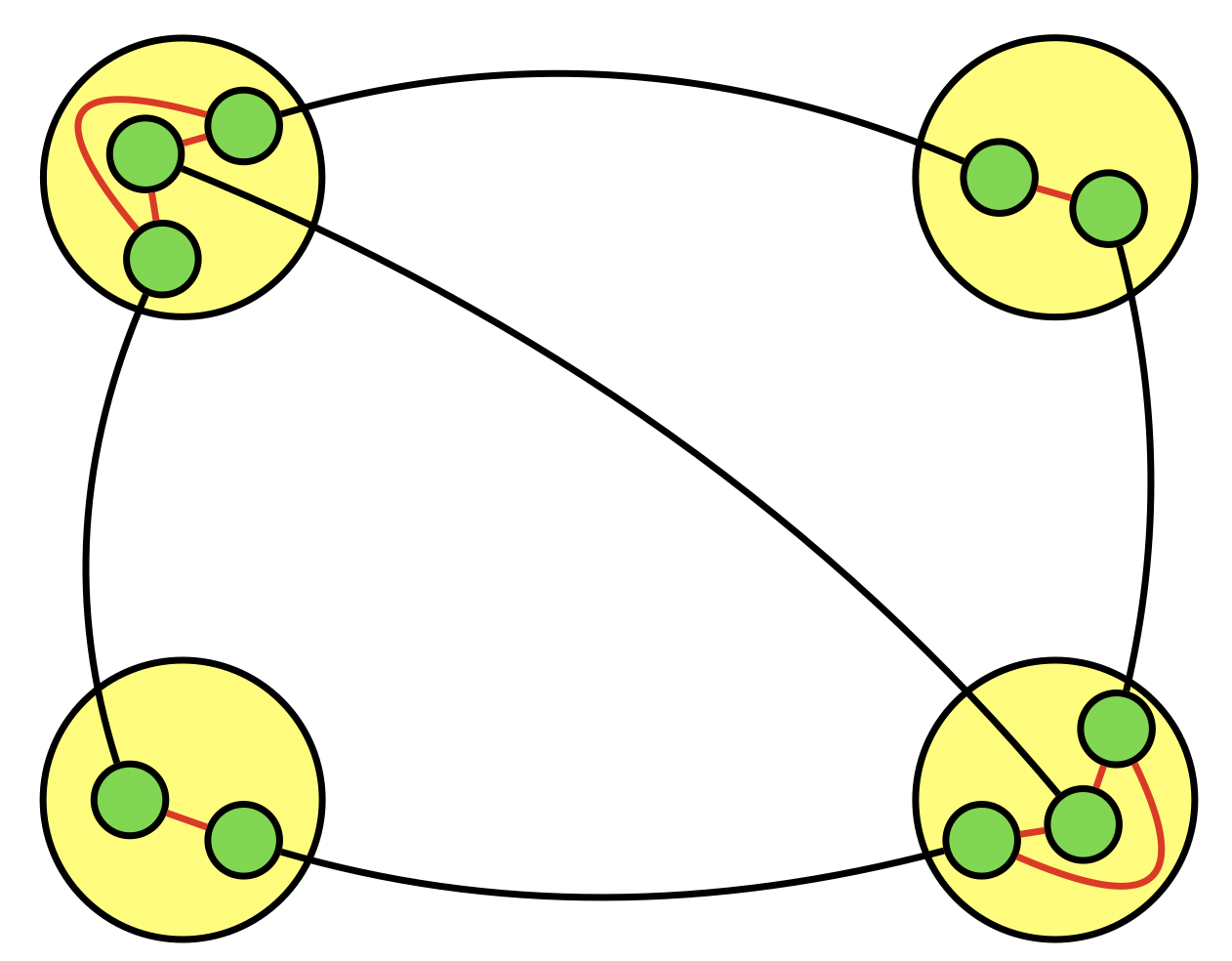}
        \caption{Graph $G$.}
                \label{subfig:sseh2}
    \end{subfigure}
    \caption{The construction of \Cref{defn:construction}. \Cref{subfig:sseh1}: a graph $H$.  \Cref{subfig:sseh2}: the graph $G$ constructed from $H$, where each vertex $v$ in $H$ is replaced by $\deg_H(v)$ green copies in $G$ and the yellow sets $V_v$ induce copies of $X_{\deg_H(v)}$, whose edges are in red.}
    \label{fig:sseh}
\end{figure}

The (poly-time) expander construction we make use of is as follows.
\begin{theorem}[Constant Degree Expander Construction \cite{hoory2006expander}]\label{thm:expander}
    There are absolute constants $d\geq3,b>0$ such that for every $n\geq1$ there is a poly-time algorithm that constructs a graph $X_n$ with $n$ vertices and maximum degree at most $d$, such that $\beta\left(X_n\right)\geq b$ whenever $n\geq2$.
\end{theorem}
\noindent We now formally describe how, given $H$, we construct $G$ so that $H$'s small set expansion relates to $G$'s spanning tree congestion.
\begin{definition}[Reduction from SSE to STC]\label{defn:construction}
    Given a graph $H$, we construct $G$ as follows.
    \begin{itemize}
        \item \textbf{Vertices.} For each $v\in V(H)$, let $V_v:=\{v_e:e\in E(H),\,v\in e\}$ contain one copy of $v$ for each incident edge. Set $V(G):=\bigcup_{v\in V(H)}V_v$.
        \item \textbf{Edges.} For each $v\in V(H)$, add edges within $V_v$ so that $G[V_v]$ is the expander $X_{\deg_H(v)}$. For each $e=\{u,v\}\in E(H)$, also add the edge $\{u_e,v_e\}$.
    \end{itemize}
\end{definition}
\noindent See \Cref{fig:sseh} for an illustration of the construction.
\noindent It is easy to see that $G$ is connected, has maximum degree at most $d+1$ (where $d$ is a constant) and has $\left|\bigcup_{v\in S}V_v\right|=\vol_H(S)$ for every $S\subseteq V(H)$.

\begin{figure}[t]
    \centering
    \begin{subfigure}[t]{0.25\textwidth}
        \centering
        \includegraphics[width=.95\linewidth]{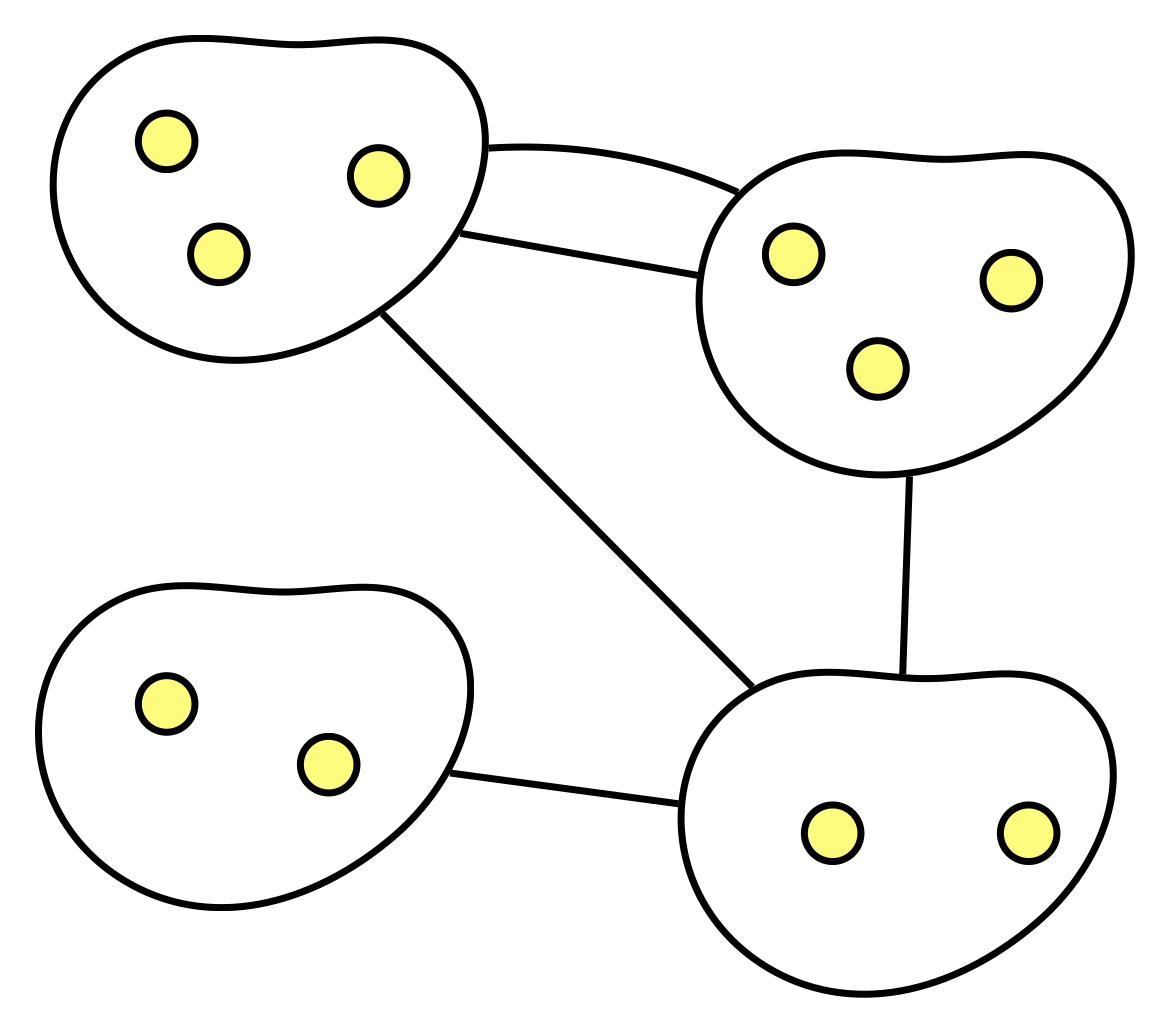}
        \caption{A balanced partition in $H$.}
                \label{subfig:ssehyeah1}
    \end{subfigure}%
    ~
    \begin{subfigure}[t]{0.25\textwidth}
        \centering
        \includegraphics[width=.95\linewidth]{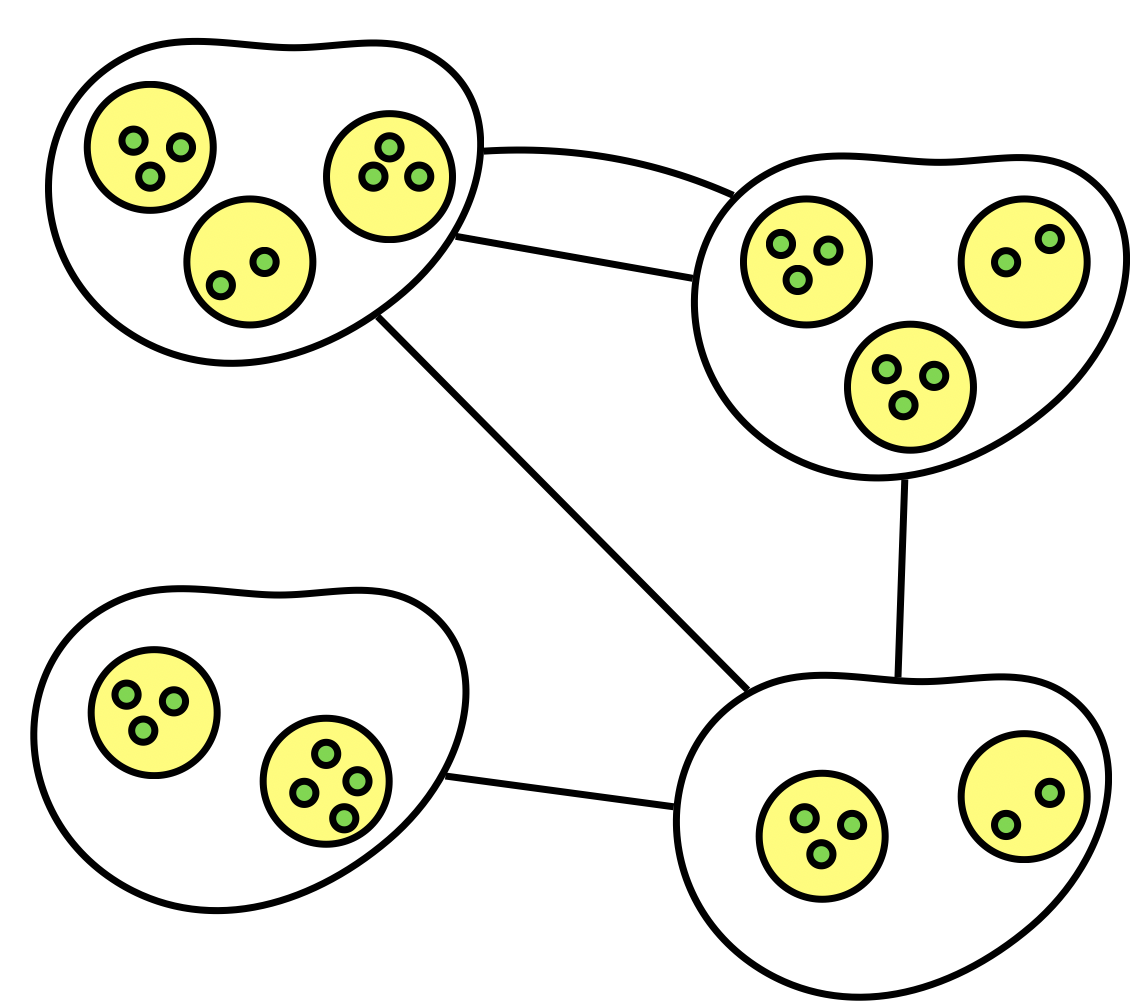}
        \caption{A balanced partition $G$.}
                \label{subfig:ssehyeah2}
    \end{subfigure}%
    ~
    \begin{subfigure}[t]{0.25\textwidth}
        \centering
        \includegraphics[width=.75\linewidth]{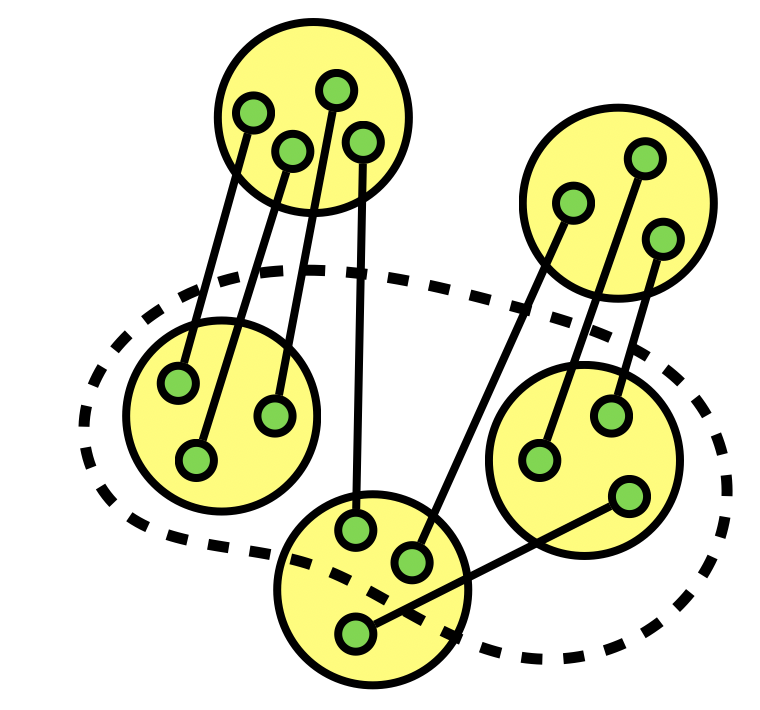}
        \caption{An expanding set in $G$.}
                \label{subfig:ssehnope1}
    \end{subfigure}%
    ~
    \begin{subfigure}[t]{0.25\textwidth}
        \centering
        \includegraphics[width=.75\linewidth]{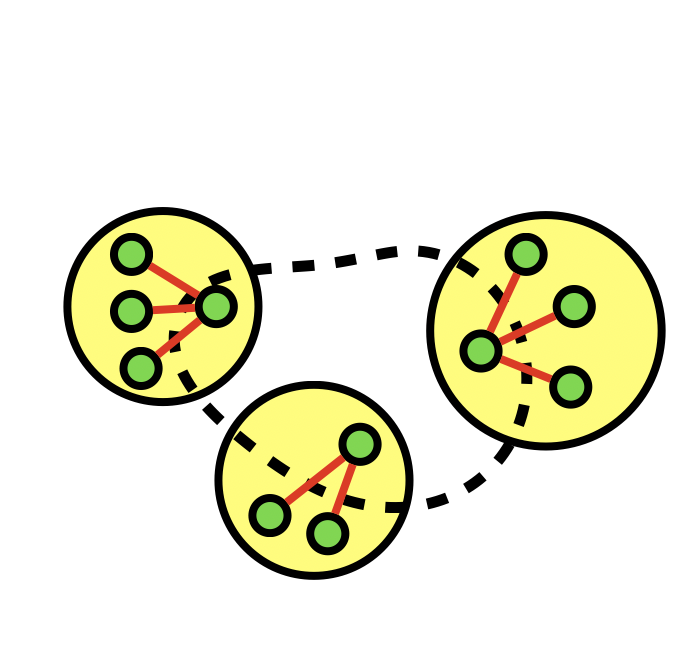}
        \caption{An expanding set in $G$.}
                \label{subfig:ssehnope2}
    \end{subfigure}
    \caption{Mapping YES/NO to YEAH/NOPE in the proof of \Cref{lemma:Gpreserves}. \Cref{subfig:ssehyeah1}: the balanced partition in $H$ in the YES case. \Cref{subfig:ssehyeah2}: the corresponding balanced partition in $G$. The same number of edges cross. \Cref{subfig:ssehnope1}: an expanding set $S$ (in a dotted circle) in $G$ that takes the majority of many $V_v$'s. In this case there are many black edges crossing the $V_v$'s that cross $S$. \Cref{subfig:ssehnope2}: an expanding set $S$ (in a dotted circle) in $G$ that is spread out across many $V_v$'s. In this case there are many red expander edges within the $V_v$'s that cross $S$.}
    \label{fig:ssehyeahnope}
\end{figure}

We begin by observing that the computational hardness of $H$ carries over to $G$ even if we additionally require that each part of our balanced partition is connected.
\begin{lemma}[$G$ preserves $H$'s YES/NO answer]\label{lemma:Gpreserves}
    Assuming the Small Set Expansion Hypothesis is true, for every sufficiently small $\eps > 0 $ it is NP-hard to distinguish whether $G$ as defined in \Cref{defn:construction} satisfies
    \begin{itemize}
        \item \textbf{YEAH (Balanced Connected Partition)}: there is a partition $W_1 \sqcup W_2 \sqcup \ldots$ of $V(G)$ such that $G[W_i]$ is connected and $|W_i|\leq\epsilon\cdot|V(G)|$ for every $i$ and $\left|\bigcup_i\delta_G(W_i)\right|\leq\epsilon\cdot|V(G)|$ or
        \item \textbf{NOPE (Balanced Sets Expand)}: every $S\subseteq V(G)$ satisfying $|V(G)|/5\leq|S|\leq4|V(G)|/5$ also satisfies $|\delta_G(S)|\geq \Omega(\sqrt{\epsilon}\cdot|V(G)|)$.
    \end{itemize}
\end{lemma}
\begin{proof}
    We show that if $H$ is YES then $G$ is YEAH, and if $H$ is NO then $G$ is NOPE. Throughout, we use the fact that the construction of $G$ takes polynomial time by \Cref{thm:expander}.

    If $H$ is YES, then let $\{V_i\}_i$ be the partition of $V(H)$ as described in \Cref{thm:sseh}. We'll start with the partition $\left\{W_i:=\bigcup_{v\in V_i}V_v\right\}_i$ of $G$ and adjust it to be a partition satisfying the conditions of YEAH. First observe that $\left|W_i\right|=\vol_H\left(V_i\right)\leq \epsilon\cdot\vol(H)=\epsilon\cdot\left|V(G)\right|$. Furthermore, by the construction there is a bijection between crossing edges of $\left\{W_i\right\}$ and crossing edges of $\left\{V_i\right\}$, for which there are at most $\epsilon\cdot\vol(H)=\epsilon\cdot \left|V(G)\right|$ by the YES assumption. Now consider the partition obtained by replacing each $W_i$ with the sets corresponding to its connected components in $G[W_i]$. The sizes of the connected components can only go down, and the number of crossing edges cannot increase. Then this is a suitable partition for YEAH.

    If $H$ is NO, then fix any $S\subseteq V(G)$ with $|V(G)|/5\leq|S|\leq4|V(G)|/5$. We need to relate the vertex copies comprising $S$ with their originals in $H$. To that end, define 
    $$ S^{maj}_H:=\left\{v\in V(H):\left|S\cap V_v\right|\geq |V_v|/2\right\}, S^{maj}_G:=\bigcup_{v\in S^{maj}_H} V_v.$$
    In other words, $S^{maj}_H$ contains the originals in $H$ for which $S$ contains at least half of the copies in $G$.
    We need to bound $\left|\delta_G(S)\right|$. 
    The idea is that either $S$ takes the majority of many $V_v$'s, in which case $\delta_G(S)$ has many crossing edges between planted expanders, or $S$ is spread out across many $V_v$'s, in which case $\delta_G(S)$ has many crossing expander edges.
    Observe that the set $S\triangle S^{maj}_G$ contains the copies of $v\in S^{maj}_H$ in $G$ that are not in $S$ and copies of $v\not\in S^{maj}_H$ in $G$ that are in $S$ and tells us exactly which case we are in.
    We show 
    \begin{equation}\label{equ:nope}
        \left|\delta_G(S)\right|\geq \max\left\{
        b\cdot\left|S\triangle S^{maj}_G\right|, \frac{b}{b+d+1}\cdot\left|\delta_H\left(S^{maj}_H\right)\right|\right\}. 
    \end{equation} 
    To prove the first bound in \eqref{equ:nope} we only need to count the edges in $\delta_G(S)$ that are strictly inside a $V_v$, that is, edges within planted expanders. By definition of $S^{maj}_G$ such edges cross $S$ exactly when they cross $S\triangle S^{maj}_G$ and we have 
    \[
        |\delta_G(S)|\geq\sum_{v\in V(H)}\left|\delta_{G[V_v]}\left((S\triangle S^{maj}_G)\cap V_v\right)\right|\geq b\cdot|S\triangle S^{maj}_G|
    \]
    by $b$-expansion.
    To prove the second bound in \eqref{equ:nope} we count edges that cross $S^{maj}_G$, that is, edges between the planted expanders, and by the construction we have $\left|\delta_G\left(S^{maj}_G\right)\right|=\left|\delta_H\left(S^{maj}_H\right)\right|$. Then we have 
    \begin{align*}
        \left|\delta_H\left(S^{maj}_H\right)\right|=\left|\delta_G\left(S^{maj}_G\right)\right| &\leq \left|\delta_G\left(S\right)\right|+\left|\delta_G\left(S\triangle S^{maj}_G\right)\right| 
        \\ &\leq \left|\delta_G\left(S\right)\right|+(d+1)\left|S\triangle S^{maj}_G\right| 
        \\&\leq \left|\delta_G\left(S\right)\right| + \left(\frac{d+1}{b}\right)\left|\delta_G\left(S\right)\right| = \left|\delta_G\left(S\right)\right|\cdot\left(1+\frac{d+1}{b}\right)
    \end{align*}
    where the first line is because an edge crossing $S^{maj}_G$ must cross $S$ or $S\triangle S^{maj}_G$, the second line is because the maximum degree of $G$ is at most $d+1$ and the third line uses the first bound in \eqref{equ:nope}. Rearranging gives the second term. 

    To actually bound the terms of \eqref{equ:nope} we case on whether $\left|S\triangle S^{maj}_G\right|$ is large or not. In the large case, $\left|S\triangle S^{maj}_G\right|>\left|V(G)\right|/10$, in which the planted expander edges alone (the first term in \eqref{equ:nope}) give $\left|\delta_G(S)\right|\geq b\cdot\left|S\triangle S^{maj}_G\right|\geq b\cdot\left|V(G)\right|/10$. Otherwise, in the small case, $\left|S\triangle S^{maj}_G\right|\leq \left|V(G)\right|/10$, so $S^{maj}_H$ satisfies 
    $$\vol_H\left(S^{maj}_H\right)=\left|S^{maj}_G\right|\in\left[\frac{\left|V(G)\right|}{5}-\frac{\left|V(G)\right|}{10},\frac{4\left|V(G)\right|}{5}+\frac{\left|V(G)\right|}{10}\right].$$
    Then the NO case tells us that $S^{maj}_H$ expands, i.e.\ $\left|\delta_G(S)\right|\geq\Omega\left(\left|\delta_H\left(S^{maj}_H\right)\right|\right)\geq\Omega\left(\sqrt{\epsilon}\cdot\left|V(G)\right|\right)$ where the first inequality uses \eqref{equ:nope} and the fact that $b,d$ are constants, and the second uses the NO guarantee and $\vol(H)=|V(G)|$.

    See also \Cref{fig:ssehyeahnope} for an illustration of how YES/NO map to YEAH/NOPE.
\end{proof}
\noindent
Concluding, the hardness of spanning tree congestion follows by the hardness of distinguishing between YEAH and NOPE as above.

\lowerbound*
\begin{proof}[Proof of \Cref{thm:hardness}]
    Fix any constant $C\geq1$. We show that a $C$-approximation for spanning tree congestion on graphs of maximum degree at most $D:=d+1$ distinguishes the NP-hard YEAH and NOPE cases of \Cref{lemma:Gpreserves}. 

    If $G$ is YEAH, let $\{W_i\}_{i\in[k]}$ be the partition of $V(G)$ as guaranteed. Choose any spanning tree $T_i$ of each $G[W_i]$ and construct a spanning tree of $G$ by arbitrarily adding edges from the set of edges crossing the partition to connect the $T_i$'s. Since $G[W_i]$ has at most $(d+1)|W_i|\leq (d+1)\epsilon\cdot|V(G)|=O(\epsilon\cdot|V(G)|)$ edges (since $G$ has maximum degree at most $d+1$ and $d$ is a constant) and there are at most $\epsilon\cdot|V(G)|$ crossing edges, we have an upper bound of 
    $ \stc(G) \leq O\left(\epsilon\cdot |V(G)|\right) $ in this case.
    
    If $G$ is NOPE, consider a minimum congestion spanning tree $T^*$ of $G$ and its centroid $r$. We can group up the components of $T^{*{-r}}$ into two groups such that each group has at least $(|V(G)|-1)/3$ vertices. Let $S$ be the union of $r$ and one of the groups; for large enough $G$ we have $|V(G)|/5\leq|S|\leq4|V(G)|/5$. Since $G$ has maximum degree at most $d+1$ we have $\deg_{T^*}(r)\leq \deg_G(r)\leq d+1$. Every edge of $\delta_G(S)$ congests in $T^*$ one of the $\deg_{T^*}(r)=O(1)$ edges of $T^*$ incident to $r$. By averaging, we have a lower bound of $\stc(G) \geq \Omega\left(\sqrt{\epsilon}\cdot|V(G)|\right)$ in this case.

    Unpacking our $O$ notation, we have that there are constants $c_1,c_2 > 0 $ for which YEAH gives $\stc(G)\leq c_1\epsilon\cdot |V(G)|$ and NOPE gives $\stc(G)\geq c_2\sqrt{\epsilon}\cdot|V(G)|$. Choose a sufficiently small constant $\epsilon>0$ such that $Cc_1\epsilon<c_2\sqrt{\epsilon}$. Then a $C$-approximation distinguishes YEAH and NOPE by comparing its output's congestion with $c_2\sqrt{\epsilon}\cdot|V(G)|$. Thus, \Cref{lemma:Gpreserves} implies that approximating spanning tree congestion within a factor of $C$ is NP-hard under the Small Set Expansion Hypothesis. 
\end{proof}

\section{Acknowledgements}
We thank Jannis Blauth for many useful conversations on this problem over the past couple of years, and Rajmohan Rajaraman for recent fruitful discussions.

\paragraph{AI Disclosure.} An initial algorithm using a more complicated decomposition and SSE hardness result were discovered by Claude Opus 5.5. The authors refined and simplified both results by simplifying the decomposition, replacing a subroutine with cutmatches, and connecting the SSEH statement to \cite{austrin2012inapproximability}. The authors wrote the manuscript. OpenAI's GPT-6 Astra was used to proofread the manuscript. The authors verified the correctness of the results and take full responsibility for them.

\bibliographystyle{alpha}
\bibliography{references}

\appendix

\section{Integrality Gap of the Natural LP for Spanning Tree Congestion}\label{sec:IG}
In this section, we observe that the natural LP for spanning tree congestion has an integrality gap of $\Omega(n)$.

For a connected graph $G=(V,E)$, this relaxation is defined as follows. Let $P_{\mathrm{ST}}(G)$ be the spanning tree polytope of $G$ (i.e.\ the convex hull of the incidence vectors $\chi^T$ of its spanning trees~\cite{edmonds1971matroids}). The relaxation minimizes $c$ subject to the constraint that the weight across every cut is at least $1/c$ times the number of graph edges crossing that cut; for each fixed $c>0$, these are LP constraints.
\begin{equation}\tag{STC LP}\label{eq:stc-lp}
    \begin{alignedat}{2}
        &\min\quad c &&\\[2pt]
        &\text{s.t.}\quad\sum_{e\in\delta_G(S)}x_e\geq \frac{|\delta_G(S)|}{c}
            &&\qquad\forall S\subseteq V\\
        &\qquad x\in P_{\mathrm{ST}}(G),\quad c>0. &&
    \end{alignedat}
\end{equation}
It is a simple exercise to verify that this relaxes minimum spanning tree congestion.

Letting $c_{\mathrm{LP}}(G)$ be the optimal value of \eqref{eq:stc-lp}, we have the following integrality gap lower bound.
\begin{theorem}[Integrality Gap]\label{thm:integralitygap}
    There is a family of connected $n$-vertex graphs $G$ for which $c_{\mathrm{LP}}(G)=O(1)$ and $\stc(G)=\Omega(n)$.
\end{theorem}
\begin{proof}
    For fixed $n$, we let $G=X_n$ as in \Cref{thm:expander}. It follows that $\Delta_G\leq d$ and $\beta(G)\geq b$ for absolute constants $d,b>0$.

    We first argue that $c_{\mathrm{LP}}(G)=O(1)$. Each edge shares an endpoint with at most $2d-2$ other edges, so a greedy edge coloring partitions $E$ into $q\leq 2d-1$ matchings $M_1,\ldots,M_q$. Since each matching is a forest and $G$ is connected, each $M_i$ can be extended to a spanning tree $T_i$. Set
    \[
        x:=\frac{1}{q}\sum_{i=1}^q\chi^{T_i}.
    \]
    By construction, $x\in P_{\mathrm{ST}}(G)$. Furthermore, every edge belongs to some $M_i\subseteq E(T_i)$, so $x_e\geq1/q$ for every $e\in E$. Consequently, $x$ with $c=q$ is feasible for \eqref{eq:stc-lp}, giving $c_{\mathrm{LP}}(G)\leq q=O(1)$.

    We next argue that $\stc(G)=\Omega(n)$. Let $T$ be any spanning tree of $G$ and let $r$ be a centroid of $T$ with respect to $V$. The components of $T^{-r}$ contain $n-1$ vertices in total. There are at most $d$ such components, each with at most $n/2$ vertices, so the vertex set $C$ of a largest component satisfies $(n-1)/d\leq |C|\leq n/2$. The unique tree edge joining $C$ to $r$ has congestion $|\delta_G(C)|$. By the expansion of $G$ we get
    \[
        \con_G(T)\geq |\delta_G(C)|\geq b|C|\geq\frac{b(n-1)}{d}=\Omega(n).
    \]
    Since this holds for every spanning tree $T$, we obtain $\stc(G)/c_{\mathrm{LP}}(G)=\Omega(n)$.
\end{proof}

\end{document}